\documentclass[a4paper,fleqn,usenatbib]{mnras}

\usepackage[T1]{fontenc}
\usepackage{ae,aecompl}
\usepackage{graphicx}	% Including figure files
\usepackage{amsmath}	% Advanced maths commands
\usepackage{amssymb}	% Extra maths symbols
\usepackage{color}
\newcommand{\lya}{Ly$\alpha$ }

\def\gtorder{\mathrel{\raise.3ex\hbox{$>$}\mkern-14mu
    \lower0.6ex\hbox{$\sim$}}} 
\def\ltorder{\mathrel{\raise.3ex\hbox{$<$}\mkern-14mu
    \lower0.6ex\hbox{$\sim$}}}
    
\title[\lya Properties of Galaxies in Overdense Regions]{\lya Properties of Simulated Galaxies in Overdense Regions: Effects of Galactic Winds at $z \gtorder 6$}

\author[R. Sadoun et al.]{
Raphael Sadoun,$^{1,2}$\thanks{E-mail: sadoun@astro-osaka.jp}
Emilio Romano-D\'{\i}az,$^{3}$
Isaac Shlosman$^{1,4}$
and Zheng Zheng$^{2}$
\\
$^{1}$Department of Earth and Space Science, Graduate School of Science, Osaka University, Osaka 560-0043, Japan\\
$^{2}$Department of Physics \& Astronomy, University of Utah, Salt Lake City, UT 84112-0830, USA\\
$^{3}$Argelander Institut fuer Astronomie, Auf dem Haegel 71, D-53121 Bonn, Germany\\
$^{4}$Department of Physics \& Astronomy, University of Kentucky, Lexington, KY 40506-0055, USA
}

\date{Accepted XXX. Received YYY; in original form ZZZ}

\pubyear{2018}

\begin{document}
\label{firstpage}
\pagerange{\pageref{firstpage}--\pageref{lastpage}}
\maketitle

% Abstract
\begin{abstract}
We perform Monte-Carlo radiative transfer calculations to model the \lya properties of galaxies in high-resolution, zoom-in cosmological simulations at $z \sim 6.6$. The simulations include both constrained and unconstrained runs, representing respectively a highly overdense region and an average field. Different galactic wind models are used in the simulations in order to investigate the effects of these winds on the apparent \lya properties of galaxies. We find that, for models including galactic winds, the \lya properties of massive galaxies residing in the overdense region match well recent observations of luminous \lya emitters (LAEs) at $z\sim 6-7$, in terms of apparent \lya luminosity, \lya line width and \lya equivalent width distributions. Without winds, the same galaxies appear less \lya bright as a result of both differences in the line profile emerging from galaxies themselves, and, in the distributions of neutral gas in the circumgalactic (CGM) and intergalactic medium (IGM). We also study the relations between apparent \lya luminosity and various galaxy properties: stellar mass, star formation rate (SFR) and host halo mass. At fixed \emph{halo mass}, the apparent \lya luminosity of galaxies appears to depend on the large-scale environment while this is no longer true for galaxies at a given \emph{stellar mass} or SFR. We provide simple linear fits to these relations that can be used for quickly constructing mock LAE samples from N-body simulations. Our results suggest that the observed luminous LAEs at $z\sim 6.6$ are hosted by $\sim 10^{12}\,h^{-1}\,\mathrm{M}_\odot$, dark matter haloes, residing in large, overdense ionized regions. 

\end{abstract}

% Keywords

%%%%%%%%%%%%%%%%%%%%%%%%%%%%%%%%%%%%%%%%%%%%%%%%%%

%%%%%%%%%%%%%%%%% BODY OF PAPER %%%%%%%%%%%%%%%%%%

%%%%%%%%%%%%%%%%%%%%%%%%%%%%%%%%%%%%%%%
\section{Introduction}
\label{sec:intro}
%%%%%%%%%%%%%%%%%%%%%%%%%%%%%%%%%%%%%%%

According to the standard model for structure formation, massive galaxies at high-redshift are found predominantly in rare, overdense regions of the Universe. Galaxy evolution is expected to proceed more rapidly in these dense environments, with elevated gas accretion and star formation rates (SFRs) that can power strong galactic outflows \citep{Romano2011b,Romano2014,Sadoun2016}. The resulting galaxy population is thus expected to exhibit significant differences in terms of physical and observational properties compared to average fields \citep[e.g.,][]{Shlosman2013}. Observationally, there are only a handful of constraints on how high-redshift galaxy formation depends on environment \citep[e.g.,][]{Dayal2018}. Bright quasars (QSOs) detected at $z \gtorder 6$, less than $\sim 1$ Gyr after the Big Bang, are thought to be formed in highly biased overdensity peaks \citep{Romano2011a,Stiavelli2005, Zheng2006,Kim2009,Maselli2009}. Yet, little is known about the properties of their host galaxies and of galaxies residing in their vicinity.

Understanding the formation of massive galaxies at high-$z$ and their dependence on environment is crucial for interpreting observations of galaxies at these early epochs. Despite the increasingly large number of galaxies detected at $z \gtorder 6$ \citep[e.g.,][]{Bouwens2014,Bouwens2015,Harikane2018,Ono2018}, galaxy surveys at these epochs still only pick up the ``tip of the iceberg'' of the galaxy population. This is mainly due to the difficult nature of detecting faint objects at high-$z$. Faint galaxies have been detected with gravitational lensing observations that can reach deeper than normal surveys \citep[e.g.,][]{Coe2015,Kawamata2016,Schmidt2016,Bouwens2017,Lotz2017,Ishigaki2018}, but the number of detected objects is usually too small for a statistical study of the overall galaxy population. 

Galaxies at high redshifts are usually detected using colour-selection techniques from broadband photometry. Galaxies detected this way are referred to as Lyman-Break Galaxies (LBGs), since they are selected on the basis that their apparent flux is suppressed blueward of the Lyman limit. An increasingly large number of high-$z$ galaxies have now also been observed as \lya emitters \citep[LAEs,][]{Kashikawa2006,Kashikawa2011,Hu2010,Ota2010,Ouchi2010,Ouchi2017,Ono2012,Caruana2014,Konno2014,Matthee2015,Santos2016,Zheng2017}. LAEs are galaxies selected from their strong \lya emission usually using narrowband imaging surveys. The importance of the \lya line and its use to detect high-$z$ galaxies were first discussed by \cite{Partridge1967}. Since then, many surveys have been able to detect LAEs at $z\gtorder 6$. In particular, the largest high redshift LAE survey to date comes from the SILVERRUSH program using Hyper Suprime Cam on the Subaru Telescope \citep{Ouchi2017}, yielding $\sim 2000$ LAEs in a $15-20\,\mathrm{deg}^2$ region at $z\sim 6-7$. Thanks to the large volume covered by the survey, SILVERRUSH has been able to probe the bright end of the \lya galaxy population at $z\sim 5.7-6.6$, reaching apparent \lya luminosities $L_{\rm Ly\alpha} \gtorder 10^{43.5-44}\,\mathrm{erg}\, \mathrm{s}^{-1}$ \citep{Konno2018,Shibuya2018}.

LAEs are thought to be young, active star-forming galaxies in which a significant fraction of the ionizing photons emitted from massive stars are converted to \lya photons as a result of HI recombination processes in the interstellar medium (ISM). After they are emitted, \lya photons escape their production site and interact with the surrounding neutral gas. The propagation of these photons out of the ISM and through the circumgalactic (CGM) and intergalactic medium (IGM) is not trivial, owing to the resonant nature of the \lya transition line. Typically, \lya photons experience a large number of scatterings, which increase their path lengths significantly compared to optically thin lines and continuum photons. The \lya flux and spectral shape that emerge out of a galaxy depend on the density, temperature, velocity distribution, and ionization state of the interstellar gas. Dust can also play a key role in modifying the radiative transfer of these photons \citep[for a comprehensive review on \lya radiative transfer, see, e.g.,][]{Dijkstra2017}. 

Once \lya photons escape from their host galaxies, they also interact with neutral hydrogen in the CGM and IGM, which further modifies the apparent \lya properties of galaxies. The \lya line therefore encodes information about the thermodynamical state of the CGM and IGM gas through which it has traveled. For this reason, LAEs offer a unique and promising tool to study the ionization state of the IGM in the late stages of the reionization era ($z\gtorder 6$; see \citealt{Dijkstra2017} for a comprehensive review of the topic). Few observational probes of the IGM exist at this period, which makes LAEs an important tracer of the late reionization history of the Universe. However, due to the resonant scattering nature of the process, extracting constraints on reionization from apparent \lya properties of LAEs is usually very challenging and requires detailed modeling of the \lya line and its transfer through the IGM gas. 

One of the main factors that can affect the radiative transfer of \lya photons in and around galaxies is the presence of galactic outflows. Early studies trying to solve analytically the \lya radiative transfer in galaxies have shown that the emergent \lya spectral line shape expected for a uniform, static gas cloud is a characteristic double peak profile centered on the systemic redshift with an offset that depends on the optical depth of the cloud \citep{Harrington1973,Neufeld1990}. In the presence of outflows however, the peak blueward of the line center is suppressed, because photons in that region appear closer to resonance in the rest frame of the outflowing gas through which photons scatter. Given that the typical \lya line profile of observed LAEs usually appear as a single peak offset from the systemic redshift, galactic outflows have often been invoked to explain the observed spectral shapes of LAEs \citep[e.g.,][]{Kunth1998,Atek2008, Hayes2015,Gronke2017}. Most notably, the so-called shell model in which \lya photons scatter in an outflowing shell of neutral gas has been very successful at reproducing many of the observed spectral features in the spectra of LAEs \citep{Ahn2004,Verhamme2006,Schaerer2011}.

Despite the recent outstanding progress in detecting LAEs at high redshift, many open questions remain on the exact nature of LAEs, their relationship to the overall galaxy population at $z\gtorder 6$, and the type of dark matter haloes where they are embedded. By obtaining constraints on the clustering length of LAEs at $z\sim6.6$, \citet{Ouchi2010} has estimated the host halo mass of typical high-$z$ LAEs to be $\sim 10^{10.5}\,\mathrm{M}_{\odot}$. However, this value might underestimate the host-halo mass for the bright end of the LAE population since bright galaxies are expected to reside in biased, large ionized regions at these redshifts. For these reasons, it is necessary to better understand the relationship between apparent \lya emission and intrinsic galaxy properties and how \lya emission from galaxies depends on environment during the reionization epoch.

The aim of this paper is to model the \lya properties of galaxies obtained from high-resolution, state-of-the-art zoom-in cosmological simulations at $z\gtorder 6$ \citep{Romano2014,Sadoun2016}. A number of previous works have been performed on modeling \lya emission in galaxy formation simulations \citep[e.g.][]{Nagamine2010,Laursen2009a,Laursen2009b,Zheng2010,Zheng2011a,Zheng2011b,Verhamme2012,Lake2015,Smith2015,Behrens2018,Kimm2018,Inoue2018,Weinberger2018,Yajima2018}. Our approach is different with respect to previous studies in that (i) we focus on a highly overdense region hosting a $\sim 10^{12}\,h^{-1}\,\mathrm{M}_\odot$ dark matter (DM) halo that collapses by $z=6$; (ii) we study how different prescriptions for galactic winds affect the resulting \lya properties of the simulated galaxies; and (iii) we compare \lya properties of galaxies in the overdense region to that of galaxies formed in an average density field to investigate their environmental dependence. The \lya properties of galaxies are obtained by post-processing the outputs of the cosmological simulations with a 3D Monte-Carlo radiative transfer code \citep{Zheng2002}.

The paper is organized as follows. We start in Section 2 by presenting the numerical methods used for our simulations of the overdense and normal regions and describe the radiative transfer model we use to calculate the \lya properties of our simulated galaxies. In Section 3, we present the global statistical properties of the resulting \lya emitting galaxy population obtained from the radiative transfer calculations. In Section 4, we discuss the implications of our results on galaxy formation and evolution at high-$z$ by comparing them with recent LAE observations. Finally, we give our conclusions in Section 5.

%%%%%%%%%%%%%%%%%%%%%%%%%%%%%%%%%%%%%%%%%%%%%%%%%%%%%
\section{Numerical methods}
\label{sec:methods}
%%%%%%%%%%%%%%%%%%%%%%%%%%%%%%%%%%%%%%%%%%%%%%%%%%%%%

\subsection{Hydrodynamical simulations of overdense and normal regions}
\label{sec:simulations}

We use the suite of high-resolution cosmological zoom-in simulations presented in \citet{Sadoun2016} (hereafter S16). The simulations are run with a modified version of the Smooth Particle Hydrodynamics (SPH) N-body code \textsc{PGADGET-3}, originally described in \citet{Springel2005}. The code includes radiative cooling by primordial and metal-enriched gas, the so-called {\it Pressure Model} for star formation \citep{Schaye2008,Choi2010,Romano2011b}, supernova feedback, a sub-resolution model for the multiphase ISM, and various recipes for galactic wind \citep{Springel2003,Choi2011}.

The S16 simulations consist of both constrained (CR) and unconstrained (UCR) runs to represent an overdense and normal region respectively. The initial conditions for the CR runs are constructed to impose the formation of a $\sim 10^{12}\,h^{-1}\,\mathrm{M}_\odot$ dark matter (DM) halo in a cubic box of $20\,h^{-1}\,\mathrm{Mpc}$ (comoving) at $z=6$ using the constrained realization method \citep[e.g.,][]{Bert1987,Hoffman1991,Weygaert1996,Romano2007,Romano2011a,Romano2011b}. This method allows us to avoid the need to simulate large volumes that would normally be required to find the rare, high-density peaks that collapse into $\sim 10^{12}\,h^{-1}\,\mathrm{M}_\odot$ haloes at $z=6$. The initial conditions for the UCR runs correspond to the same parent random realization of the density field used to construct the CR runs but without the imposed constraint. 

The simulations are run from $z=199$ to $z=6$ with vacuum boundary conditions. The multimass approach is used to downgrade the numerical resolution outside the central region with three resolution levels . The highest level has an effective resolution of $2\times\,512^3$, translating to a mass of $3.73\times 10^6\,h^{-1}\,{\rm M_\odot}$ and $8.88\times 10^5\,h^{-1}\,{\rm M_\odot}$ for DM and SPH particles respectively. The gravitational softening is $\epsilon_{\rm grav}= 140\,h^{-1}{\rm pc}$ (comoving), which is about $20\,h^{-1}{\rm pc}$ in 
physical units at $z=6$. We assumed the $\Lambda$CDM cosmology with WMAP5 parameters \citep{Dunkley2009}, 
$\Omega_{\rm m}=0.28$, 
$\Omega_\Lambda=0.72$, $\Omega_{\rm b}=0.045$, and $h=0.701$, where $h$ is the Hubble 
constant in units of 
$100\,{\rm km\,s^{-1}\,Mpc^{-1}}$. 

Halos and galaxies have been identified using the HOP group finding algorithm \citep{Eisenstein1998}. The volume in which we identify haloes and galaxies corresponds to an inner spherical region of (comoving) radius $2.7\,h^{-1}{\rm Mpc}$ and $4\,h^{-1}{\rm Mpc}$ for the CR and UCR simulation runs, respectively. In this paper, we consider only the simulation output at $z=6.6$, since the \lya frequency at this redshift corresponds to one of the few observation windows through the atmosphere and is, thus, a commonly targeted redshift of narrowband imaging surveys of LAEs during the reionization epoch \citep[e.g][]{Ouchi2010,Ouchi2017}.  

For the CR simulations, we run three different prescriptions for galactic winds: constant velocity wind \citep[CW,][]{Springel2003}, variable wind \citep[VW,][]{Choi2011}, and no-wind (NW) models. For the UCR simulation, we have used the constant velocity wind model. Throughout the paper, we use the names CW, VW and NW  for the CR simulations and UCW for the UCR simulation. We refer the reader to S16 for a more comprehensive reading of the set-up of our simulation suite. However, as it is the focus of the present study, we provide here a brief description of the different galactic wind prescriptions used in our simulations.

%%%%%%%%%%%%%%%%%%%%%%%%%%%%%%%%%%%%%%%%%%%%%%%%%%%%%%%%%%%%%
%\subsubsection{Wind models}
%\label{sec:winds}

\paragraph*{Constant Wind model (CW):}
In this model, winds are triggered by modifying the behavior of SPH particles and turning them into `wind' particles. Wind particles are not subject to hydrodynamical forces after experiencing the initial kick from supernovae. Wind particles are launched at a fixed constant velocity, $v_{\rm w} = 484\,{\rm km\,s^{-1}}$. The mass-loading factor of the wind, defined as $\beta_{\rm w} \equiv\dot M_{\rm w}/\dot M_{\rm SF}$, where $\dot M_{\rm w}$ is the mass loss in the wind and $\dot M_{\rm SF}$ is the SFR, is also chosen to be fixed to a value $\beta_{\rm w} = 2$ \citep[e.g.,][]{Choi2011}.

\paragraph*{Variable Wind model (VW):}
In this model, wind velocities $v_{\rm w}$ and mass-loading factors $\beta_{\rm w}$ are parametrized with respect to the stellar mass and SFR of the host galaxy. Such relations have been empirically constrained from observations of starburst galaxies at lower redshift. This model corresponds to the 1.5ME wind model of \citet{Choi2011}.

\paragraph*{No-Wind model (NW):}
This model does not implement kinematic feedback but only includes thermal feedback by supernovae.

%%%%%%%%%%%%%%%%%%%%%%%%%%%%%%%%%%%%%%%%%%%%%%%%%%%
\subsection{Grid interpolation}
\label{sec:sph_to_grid}

In order to model \lya emission from galaxies in our simulations, we first need to obtain a gridded representation of the density, temperature and velocity fields of the gas. This is necessary because both the self-shielding and \lya radiative transfer calculations, described in the following sections, are performed on a uniform Cartesian grid. Here, we describe how we interpolate the SPH particle data onto a grid, taking into account the finite volume of SPH particles. 

We have followed a similar method as described in \citet{Kollmeier2010}. The density, temperature and velocity of the gas in a given grid cell are given by  
\begin{equation}
S_{\rm cell} = \sum_{i=1}^{N_{\rm SPH}} \frac{m_iS_i}{\rho_i}\,K(r_i,h_i),
\label{eq:sph2grid}
\end{equation}
where $S_i$ indicates the physical quantity of interest (gas density, temperature, or one of the three velocity components, $v_x$, $v_y$ or $v_z$), $m_i$ is the mass of the SPH particle, and $\rho_i$ is the SPH density of the particle. $K(r_i,h_i)$ is the SPH kernel \citep{Springel2005} at a distance $r_i$ between particle $i$ to the center of the cell and $h_i$ is the SPH smoothing length of particle $i$. The sum is over all particles whose SPH kernel overlaps with the cell.

Throughout this paper, we use the fiducial value of $N_{\rm grid} = 512$ for the size of the grid in order to match the native resolution of the S16 simulations, but we also tested using a more refined grid with $N_{\rm grid} = 1024$ and found that our results on the \lya properties of galaxies were not affected (see Appendix). Furthermore, we only consider the volume corresponding to the highest resolution level in our simulations, which means that the grid covers a cubic volume of $(6\,h^{-1}\,\mathrm{Mpc})^3$ and $(8\,h^{-1}\,\mathrm{Mpc})^3$ for the CR and UCR simulations respectively.

%%%%%%%%%%%%%%%%%%%%%%%%%%%%%%%%%%%%%%%%%%%%%%%%%
\subsection{Self-shielding calculations}
\label{sec:self_shielding}

Next, we need to know the ionization state of the gas on the grid because the radiative transfer of the \lya line emitted from galaxies depends on the properties of the \emph{neutral} gas in the CGM and IGM. The S16 simulations do not include on-the-fly radiative transfer of ionizing photons. For this reason, we calculate \emph{a posteriori} the HI fraction, $\mathrm{f}_{\rm HI} \equiv n_{\rm HI}/n_{\rm H}$, in each cell, resulting from both photoionization and collisional ionization. We describe the photoionization calculation in more details below. Given that our simulated volume is comparable with the typical size of ionizing HII bubbles expected at $z \sim 6-7$ \citep[e.g.,][]{Lin2016,Giri2018}, we need to account not only for photoionization by the external UV background (UVB) originated from sources located on scales larger than the simulation box, but also for the ionizing flux emitted by star-forming galaxies within the box.  

The UVB is modeled as a uniform ionizing radiation field whose intensity as a function of frequency and redshift is given by the UV background model from \citet{FG2009}. The HI fraction in a given cell is calculated from the photoionization equilibrium equation
\begin{equation}
\Gamma_{\rm UVB}\mathrm{f}_{\rm HI} = \alpha(T) (1-\mathrm{f}_{\rm HI})^2,
\label{eq:photoion_equi}
\end{equation}
where $\Gamma_{\rm UVB}$ is the photoionoization rate at the position of the cell due to the UVB and $\alpha(T)$ is the case-B recombination coefficient at temperature $T$. 

We explicitly take into account self-shielding of the gas using the iterative procedure described in \citet{Kollmeier2010}. In each cell, the neutral fraction is first initialized to $\mathrm{f}_{\rm HI} \ll 1$. Then, the ionizing optical depth from outside the box to the center of the cell is calculated along the 6 principal Cartesian directions. We use these optical depths to compute the attenuated photoionization rate along each of the 6 Cartesian directions. The average of these values is used as the mean attenuated photoionization rate $\Gamma_{\rm UVB}$ appearing in Eq. \ref{eq:photoion_equi}. This new value for $\Gamma_{\rm UVB}$ implies a different solution for f$_{\rm HI}$ in each cell using Eq. \ref{eq:photoion_equi} which, in turn, changes the ionizing optical depth. 
The process is repeated until the value of f$_{\rm HI}$ has converged to a fractional accuracy better than $10^{-2}$ everywhere on the grid.

On the other hand, the ionizing flux $Q_{\rm gal}$ emitted from a galaxy in the simulation is assumed to be proportional to its star formation rate (SFR) \citep{Schaerer2003}
\begin{equation}
Q_{\rm gal} = 10^{53.8} \frac{\mathrm{SFR}}{1 \mathrm{M}_\odot \, \mathrm{yr}^{-1}}\; (\mathrm{photons}\, \mathrm{s}^{-1}).
\label{eq:Q_gal}
\end{equation}
The unattenuated photoionization rate at a physical distance $r$ from the galaxy is calculated as 
\begin{equation}
\Gamma_{\rm gal} = \frac{f_{\rm esc}Q_{\rm gal}}{4\pi r^2},
\label{eq:Gamma_gal}
\end{equation}
where $f_{\rm esc}$ is the escape fraction of ionizing photons. In our fiducial model, we use the scaling relation between $f_{\rm esc}$ and galaxy mass, $M_{\rm gal}$ from \citet{Yajima2015}, which was obtained by modeling the continuum properties of galaxies in an overdense region using a suite of simulations similar to the one in S16. 

We do not apply any attenuation to the ionizing flux from galaxies as we did for the UVB because computing the ionizing optical depth from every galaxies to the position of a given cell would be too computationally expensive to repeat for all cells on the grid. This optically thin approximation should be valid for cells located inside HII bubbles that form around galaxies since the residual neutral fractions there are negligible. For this reason, we first calculate the neutral fractions everywhere on the grid due to photoionization by the UVB according to the method described above. Second, we compute separately the neutral fractions due to photoionization by galaxies that are obtained by solving Eq. \ref{eq:photoion_equi} with $\Gamma_{\rm UVB}$ replaced by $\Gamma_{\rm gal}$. Finally, the neutral fraction value assigned to a given cell is chosen to be the lowest of the two solutions. Doing so ensures that cells located close to a star-forming galaxy are ionized by the local ionizing flux produced by the galaxy even if the UVB is not able to penetrate in that region.

\subsection{\lya radiative transfer}
\label{sec:lyaRT}

\begin{figure*}
\centering
\includegraphics[width=0.9\linewidth]{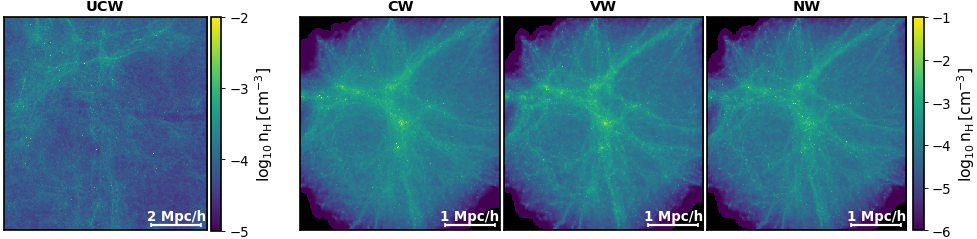}
\includegraphics[width=0.9\linewidth]{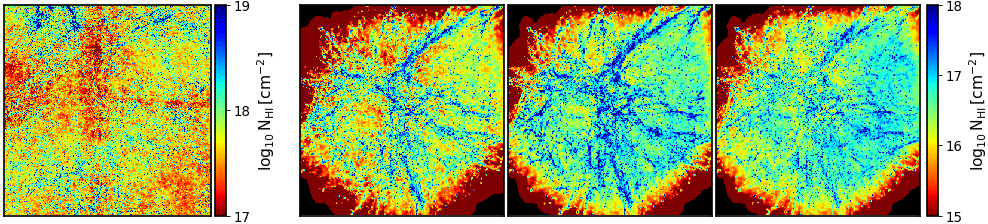}
\includegraphics[width=0.9\linewidth]{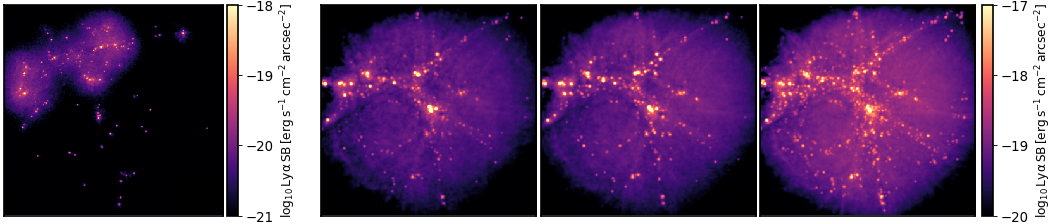}
\caption{Maps of gas distributions and \lya emission in the different simulation runs: unconstrained simulation with constant wind model (UCW) and constrained simulations with constant wind (CW), variable wind (VW) and no-wind (NW) models for galactic outflows. Top row: Gas density obtained from the interpolation of the SPH particle distribution onto a uniform grid using the SPH kernel (Section \ref{sec:sph_to_grid}). The maps show the volume-averaged density along the line-of-sight. Middle row: HI column density. The HI content is obtained from the self-shielding calculations described in Section \ref{sec:self_shielding}. Bottom row: \lya surface brightness maps resulting from our Monte-Carlo \lya radiative transfer modeling (Section \ref{sec:lyaRT}).}
\label{fig:projection_maps}
\end{figure*}

The \lya emission from our simulated galaxies is modeled using \lya radiative transfer calculations performed with a modified parallel version of the Monte-Carlo code from \citet{Zheng2002}. The code numerically follows the trajectories of \lya photons emitted by galaxies, which are treated as point sources for \lya emission in our model since we do not resolve the \lya radiative transfer in the ISM and only consider scattering in the CGM and IGM. For a given galaxy, the initial frequencies of the photons are drawn from the \lya line profile that emerge from the ISM of the galaxy. We model the profile as a simple Gaussian centered on the \lya frequency with spectral width corresponding to the Doppler broadening by the gas velocity in the cell where the galaxy is located.

Photons travel in a random direction until they are scattered. The location of a scattering event is obtained by drawing a random number $\tau$ from an exponential distribution and computing the position where the \lya optical depth is equal to $\tau$ along the propagation direction of the photon. When a photon is scattered, its frequency does not change and its new propagation direction is randomly chosen \emph{in the rest frame of the hydrogen atom}. The photon will have a different frequency and propagation direction in the lab frame that depend on the velocity and temperature of the gas at this location. Photons continue to scatter until they diffuse far enough both spatially and in frequency, and we stop following them after they escape the grid. We refer the reader to \citet{Zheng2002} for more details on the \lya radiative transfer code.

The output of the \lya radiative transfer calculation is saved in a 3-dimensional data cube containing the \lya spectrum at each pixel position, similar to an Integral Field Unit (IFU) data cube. The first two dimensions of the data cube correspond to the two spatial directions transverse to the line-of-sight and the third dimension is the spectral direction along the line-of-sight. The number of pixels in the spatial direction is $512$ in order to match the resolution of the grid while the size of a pixel in the spectral direction has been set to $0.1$\AA\ (restframe). All the \lya radiative transfer calculations presented in the paper have been run with the line-of-sight chosen along the $z$-axis of the simulation. Each time a photon scatters, we compute the probability that it escapes along the line-of-sight direction and add this probability value to the cell of the data cube corresponding to the frequency and spatial position of the photon.

In the Monte-Carlo method, the number of photons $N_\gamma$ launched from a given source is arbitrary and each ``photon'' represents a single packet carrying a fraction of the total \lya luminosity of the source. In our model, this is done by assigning a weight $w_\gamma = L_{\rm Ly\alpha}^{\rm int}/N_\gamma$ to each photon emitted 
from a source of intrinsic \lya luminosity $L_{\rm Ly\alpha}^{\rm int}$. We have found that a minimum value of $N_\gamma \sim 10^{4} \times \mathrm{SFR} /(\mathrm{M}_\odot\, \mathrm{yr}^{-1})$ is needed to properly resolve the spatial and spectral distribution of \lya emission in our simulation, and we choose $N_\gamma = 3 \times 10^{4}\times \mathrm{SFR} /(\mathrm{M}_\odot\, \mathrm{yr}^{-1})$ in our fiducial radiative transfer model.

Following \citet{Zheng2010}, we also relate the \emph{intrinsic} Ly$\alpha$ luminosity of our 
simulated galaxies to their star-formation rate, 
\begin{equation}
L_{\mathrm{Ly} \alpha}^{\rm int} = 10^{42} [\mathrm{SFR}/(\mathrm{M}_\odot \mathrm{yr}^{-1})]\, \mathrm{erg}\, \mathrm{s}^{-1},
\label{eq:lya_lum}
\end{equation}
where the SFR of each galaxy is directly obtained from the simulations. 
Note however that the results of the radiative transfer calculations, in terms of the predicted spatial and frequency distribution of 
the photons, do not depend on the value of the conversion factor between SFR and $L_{\mathrm{Ly}\alpha}^{\rm int}$. This factor is used only to convert the raw output of our radiative transfer calculations to absolute \lya flux in physical units. However the absolute \lya flux values that we predict depend on the assumed SFR-$L_{\mathrm{Ly}\alpha}^{\rm int}$ relation from Eq. \ref{eq:lya_lum}. This should be considered as one of the main uncertainties in our \lya radiative transfer model.

\begin{figure*}
\centering
\includegraphics[width=0.9\linewidth]{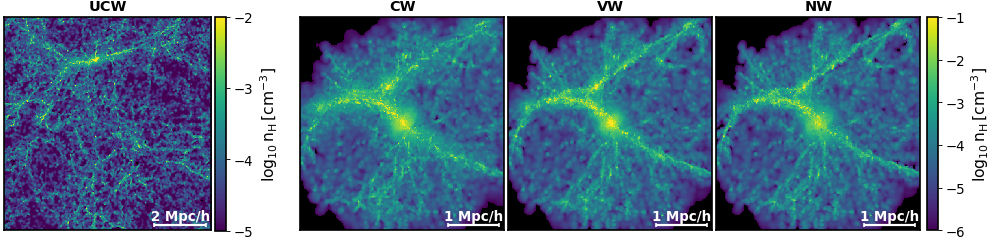}
\includegraphics[width=0.9\linewidth]{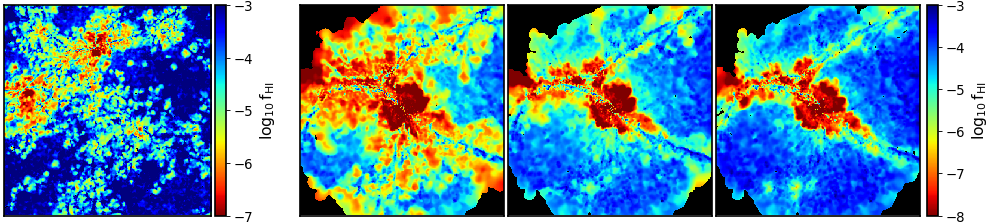}
\includegraphics[width=0.9\linewidth]{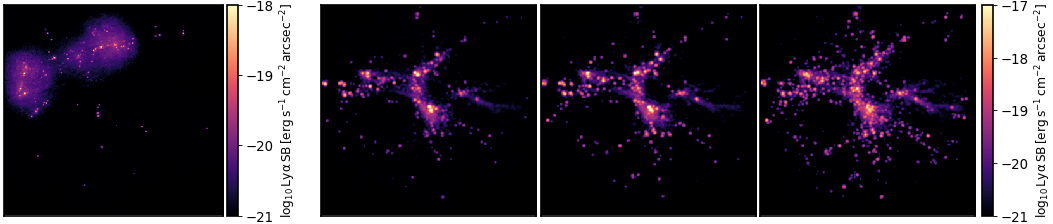}
\caption{Same as Fig. \ref{fig:projection_maps} except that the maps correspond to a single slice at the location of the most massive galaxy in each model. The middle row shows directly the values of the HI fraction $f_{\rm HI}$ instead of the HI column density as shown in Fig. \ref{fig:projection_maps}. The thickness of the slice corresponds to a single cell in the grid ($\sim 11\,h^{-1}\,{\rm comoving\, kpc}$).}
\label{fig:slice_maps}
\end{figure*}

%%%%%%%%%%%%%%%%%%%%%%%%%%%%%%%%%%%%%%%%%%%%%%%%%%%%%%%%%%%%%%%%%%%%%%%%%%%%%%%%%%
\section{Results}
\label{sec:results}
%%%%%%%%%%%%%%%%%%%%%%%%%%%%%%%%%%%%%%%%%%%%%%%%%%%%%%%%%%%%%%%%%%%%%%%%%%%%%%%%%%

In this section, we present the results of our radiative transfer calculations applied to both CR and UCR simulations. We first examine the global gas distributions and \lya emission on large scales to gain some insight on how they are affected by the different wind prescriptions and the environment. We then focus on the most massive galaxy and produce mock \lya images of the galaxy and its close environment. Next, we investigate the relation between apparent \lya luminosity and galaxy properties, namely SFR and stellar mass. 
Finally, we study the relation between \lya luminosity and host halo mass. 

\subsection{\lya view of an overdense region at $z=6.6$}
\label{sec:lya_global_view}

\begin{figure*}
\centering
\includegraphics[width=0.8\linewidth]{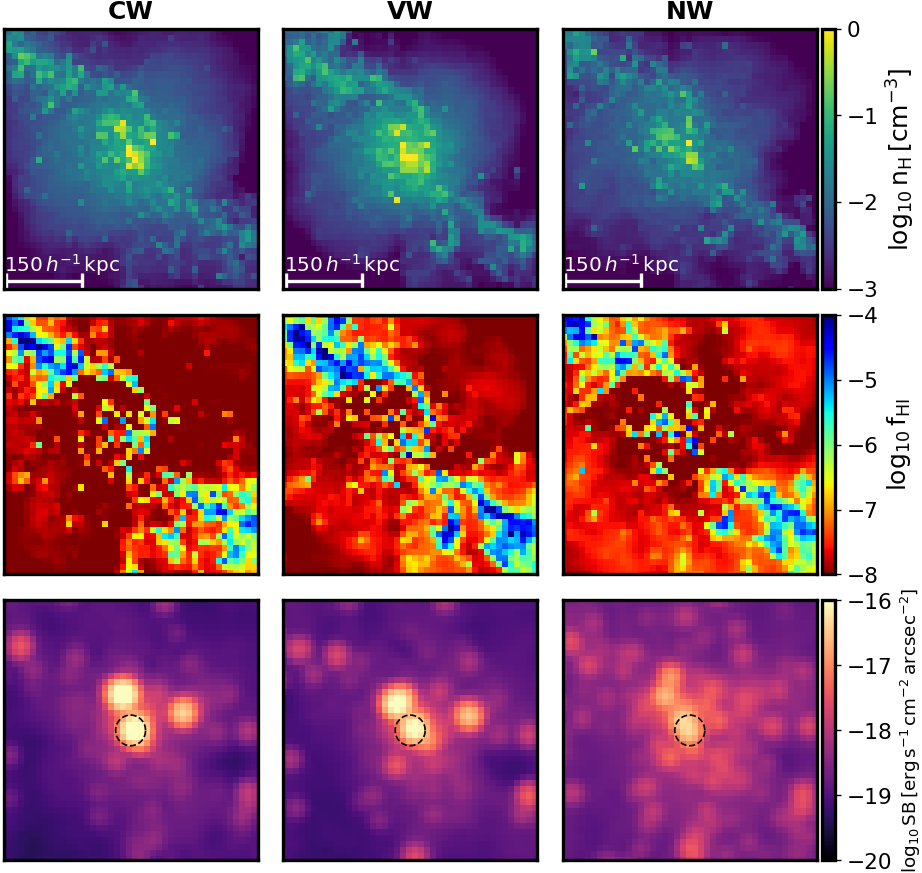}
\caption{Gas distribution around the most massive galaxy in each of the CR simulation runs (CW, VW, and NW). The maps show the total gas density (\emph{top}), neutral fraction (\emph{middle}) and mock \lya image (\emph{bottom}) in a $0.5 h^{-1}{\rm Mpc}$ region around the galaxy. The dashed circle in the \lya images indicates the 1\arcsec\ circular aperture around the galaxy which is used to calculate the apparent \lya luminosity of the source (Section \ref{sec:lya_lum}).}
\label{fig:maps_gal0}
\end{figure*}

We start by comparing the global gas distributions in our four simulation runs: the unconstrained simulation (UCW) and the three CR simulation runs with different wind prescriptions (CW, VW, and NW). Fig. \ref{fig:projection_maps} shows the maps of our simulation suite in terms of total gas density (top row, volume-average along the line-of-sight), HI column density (middle row) and \lya surface brightness (bottom row).

The gas density distributions in the CR simulations reveal a large-scale structure including dense filaments associated with the formation of the massive DM halo in the central region. The massive object is visible as a density peak in the center of the gas distribution. Overall, the gas density traces well the large-scale structure (see Fig. 2 in S16 for the maps of the DM distribution in the CR simulations) but we do find small differences between the wind models as the gas appears slightly more confined within structures for NW compared to the other two models. These differences in the distribution of baryons on large scales were already noticed and discussed in S16. We note that the position of \lya sources trace the large-scale structure as expected and that there are many more star-forming galaxies in NW compared to CW and VW as a result of insufficient feedback (see S16). 

The HI column density maps (Fig. \ref{fig:projection_maps}, middle row) show that the neutral gas mostly trace the large-scale structures as well. The highest HI column densities are associated with the dense filaments and reach values of $N_{\rm HI} \sim 10^{17}-10^{18}\,\mathrm{cm}^{-2}$. The \lya maps (Fig. \ref{fig:projection_maps}, bottom row) have been calculated using the IFU-like data cube resulting from our \lya radiative transfer calculations (Section \ref{sec:lyaRT}), by integrating over the entire datacube along the line-of-sight direction. Note that we can also produce a surface brightness map corresponding to a slice of arbitrary thickness. For example, Fig. \ref{fig:slice_maps} shows (bottom row) the \lya maps corresponding to a slice with the thickness of a single spectral pixel at the position of the most massive galaxy. The figure also displays the gas density (top row) and HI fraction distribution (middle row) in the same slice for comparison.

The HI fraction distributions in Fig. \ref{fig:slice_maps} show that the massive, central object that forms in the CR simulations reside in a large ionized bubble on a scale of $\gtorder 1\,h^{-1}\,\mathrm{Mpc}$. The spatial extent of this bubble seems to vary greatly between wind models as we observe an extended region of ionized gas with $\mathrm{f}_{\rm HI} \leq 10^{-5}$ that is present in CW but not in the other two runs. The volume occupied by the ionized region with gas of $\mathrm{f}_{\rm HI} \leq 10^{-5}$  is found to be $\sim 44\,(h^{-1}\, \mathrm{Mpc})^3$ (CW), $\sim 15\,(h^{-1}\, \mathrm{Mpc})^3$ (VW), and $\sim 11\,(h^{-1}\, \mathrm{Mpc})^3$ (NW), respectively.

From the \lya surface brightness maps in Fig. \ref{fig:projection_maps} and Fig. \ref{fig:slice_maps}, we find some variations in the spatial distribution and intensity of the \lya emission between the different models, which illustrates the effect of galactic winds on the apparent \lya emission on large-scales. While the position of the largest emission peaks with surface brightness values~$\gtrsim 10^{-18}-10^{-17}\, {\rm erg\,s^{-1}\,cm^{-2}\, arcsec^{-2}}$ (corresponding to the location of the \lya sources) remain fairly similar, the spatial extent of the diffuse 
\lya component (\lya halos) around them is different. In the NW model, this diffuse \lya emission traces relatively well the filamentary large-scale structure. However, this emission is more extended than in the other two wind models. 
We stress that the surface brightness levels shown on Fig. \ref{fig:projection_maps} and Fig. \ref{fig:slice_maps} depend on the normalization of the SFR-$L_{\mathrm{Ly}\alpha}^{\rm int}$ relation (Eq. \ref{eq:lya_lum}) but the spatial distribution of the \lya emission is not affected by this choice.

The more extended \lya emission in the NW model appears to be caused by the combination of a larger number of sources as well a larger amount of neutral gas in the simulation. In addition, the absence of galactic winds in the NW model result in \lya photons being launched closer to line center which can increase their spatial and frequency diffusion with respect to the CW and VW models that include outflows. The existence of an extended \lya emission around sources has already been predicted by radiative transfer calculations \citep[e.g.,][]{Zheng2002,Zheng2010} and later been observed by stacking imaging data of LAEs \citep[e.g.,][]{Steidel2011} or around individual LAEs \citep{Leclercq2017}. However, most narrowband imaging surveys that are designed to detect LAEs at high-redshift have a finite aperture size that can potentially prevent them to sufficiently resolve this diffuse \lya component \citep{Sadoun2017}.

%%%%%%%%%%%%%%%%%%%%%%%%%%%%
\subsection{Gas distributions and \lya emission around the most massive galaxy}
\label{sec:maingal}

We now study in more detail the gas distributions and the corresponding \lya emission around the central, most massive galaxy in each CR simulation runs. We also consider the most massive galaxy in the UCW run and compare it to its counterpart in the CW simulation. 

Fig. \ref{fig:maps_gal0} shows the gas density maps (\emph{top row}), HI fraction maps (\emph{middle row}), and \lya surface brightness maps (\emph{bottom row}) in a $500\,h^{-1}\,\mathrm{kpc}$ region centered around the most massive galaxy in each of the CR simulation runs. The gas density maps and HI fraction maps correspond to a slice of one grid cell located at the position of the galaxy along the line-of-sight. Each \lya image is produced with a spectral thickness of $131$\AA\ in the observed frame, centered on the main galaxy, to match that of the narrowband filter \emph{NB921} of Hyper Suprime Cam (HSC; e.g., \citealt{Ouchi2017,Konno2018}), and it is then smoothed with a 2D Gaussian with a full width at half-maximum (FWHM) of $1\arcsec$.

Both the gas density and HI fraction maps show the gas being accreted along a dense filament, which feeds it onto the main galaxy. The filamentary gas has a higher neutral fraction than the surrounding environment, with $\mathrm{f}_{\rm HI} \gtorder\, 10^{-4}$, compared to $\mathrm{f}_{\rm HI} \ltorder 10^{-7}$ for the material outside.  The \lya maps reveal the differences in \lya surface brightness distribution in the environment of the most massive galaxy. We see again on these maps that the diffuse \lya emission around the galaxy appears more extended in the NW run than in the other models. 

We mentioned already the issue of having a finite aperture size in resolving this diffuse emission. This is better illustrated here for the case of a 1\arcsec\ circular aperture (shown as dashed lines on the \lya maps), which is the aperture we use to define the apparent \lya luminosity of galaxies in our simulations (more details on how we calculate the apparent luminosity of galaxies are given in Section \ref{sec:lya_lum}). The maps show that the \lya surface brightness within the aperture is reduced in NW compared to CW and VW models as a result of the larger spatial diffusion.  This can partially explain why the main galaxy appears fainter in the NW model, although the main factor is probably that the galaxy has a lower intrinsic \lya luminosity (due to its lower SFR) compared to the CW and VW models. We note that there are few sources close to the central galaxy which can also contribute to the \lya flux within the aperture.

It is worth mentioning that the distribution of neutral gas around the central massive halo in the CR simulations appear highly anisotropic (Fig. \ref{fig:maps_gal0}). The apparent \lya properties of a source in an anisotropic gas distribution have been shown to strongly depend on the viewing direction. This picture has been proposed to explain the wealth of different \lya line profiles observed in the spectra of high-$z$ LAEs \citep{Zheng2014}. For this reason, the \lya luminosity of our simulated galaxies could depend on our (arbitrary) choice of a particular line-of-sight used in the radiative transfer calculations. However, the dependence of \lya properties on viewing angle should cancel out when averaged over the entire galaxy population. For this reason, we expect that our results on the statistical relation between \lya emission and galaxy properties presented in the following sections will not be altered significantly if we choose a different line-of-sight direction.

\begin{figure}
\centering
\includegraphics[width=\linewidth]{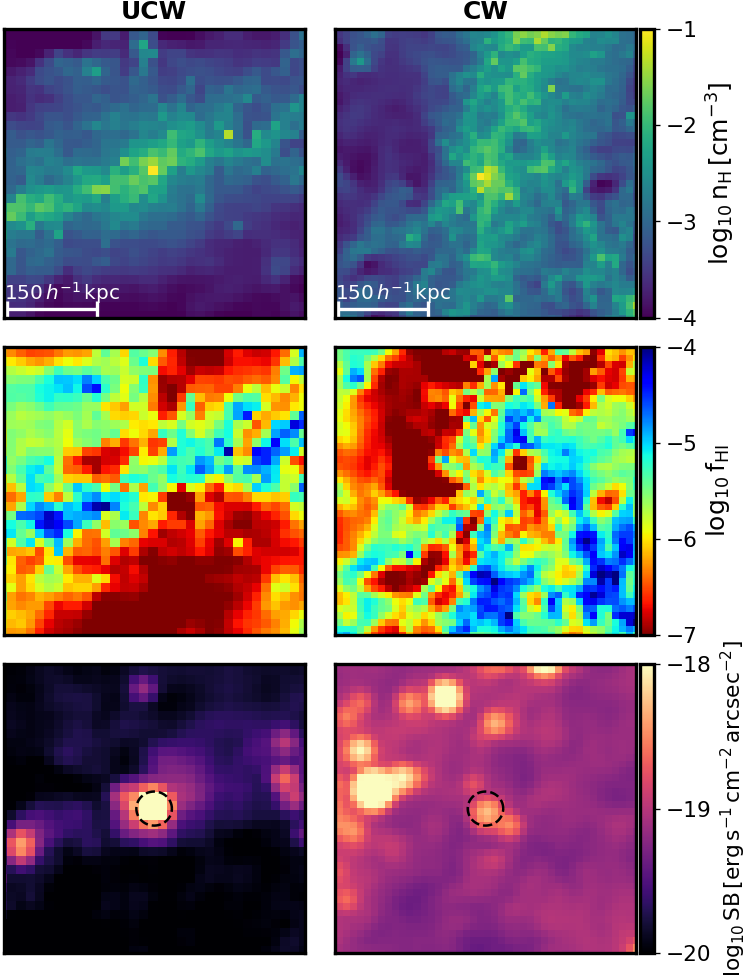}
\caption{Maps of the gas distributions around the most massive galaxy formed in the UCW simulation (\emph{left column}) and around its counterpart in the CW simulation (\emph{right column}). The maps show the total gas density (\emph{top}), neutral fraction (\emph{middle}) and \lya surface brightness (\emph{bottom}) in a $0.5\,h^{-1}\,\mathrm{Mpc}$ region around the galaxy.}
\label{fig:maps_ucwgal0}
\end{figure}

In Fig. \ref{fig:maps_ucwgal0}, we show maps of the gas density, neutral fraction and \lya surface brightness in a $0.5\,h^{-1}\,\mathrm{Mpc}$ region around the most massive galaxy in the UCW simulation (left column). The galaxy has a stellar mass of $M_\star = 1.5\times\,10^9\,h^{-1}\,\mathrm{M}_\odot$ and $\mathrm{SFR}=9.6\,{\rm M_\odot\,yr^{-1}}$, and is located in the dense filament that can be seen in the upper region of the density maps in Fig. \ref{fig:projection_maps} and Fig. \ref{fig:slice_maps}. Since the CR and UCR simulations share the same parent realization of the initial density field, we can identify, in the CW model, the counterpart of both the galaxy and the filament present in the UCW run. In the CW model, this galaxy has a stellar mass of $M_\star = 6.3\times\,10^8\,h^{-1}\,\mathrm{M}_\odot$ and $\mathrm{SFR}=4.9\,{\rm M_\odot\,yr^{-1}}$. It is located outside of the central, massive DM halo that forms in the CW simulation. The maps showing the region around this galaxy are displayed in the right column in Fig. \ref{fig:maps_ucwgal0}. 

Comparing the maps of the gas distributions in the UCW and CW runs shows that the imposed constraint affects significantly the environment around the galaxy. The filament in which the galaxy is embedded appears denser, thicker and more neutral in the CW model than in the UCW model. The \lya images show that the \lya surface brightness within the 1\arcsec\ aperture reaches $\gtorder 10^{-18}\,{\rm erg\,s^{-1}\,cm^{-2}\,arcsec^{-2}}$ in the UCW model, roughly $\sim 5$ times higher than in the CW model. This discrepancy cannot be entirely attributed to differences in the intrinsic \lya luminosity of the galaxy between CW and UCW models since their SFR only differ by a factor of $2$. Our results thus demonstrate that the large-scale environment played an important role in determining the final properties of these galaxies and of their surrounding CGM at $z\sim 6.6$, which ultimately leads to variations in their apparent \lya luminosity.

%%%%%%%%%%%%%%%%%%%%%%%%%%%%%%%%%%%%%%%%%%%%%%%%%
\subsection{Relations between apparent \lya luminosity and galaxy properties}
\label{sec:lya_lum}

\begin{figure*}
\centering
\includegraphics[width=0.9\linewidth]{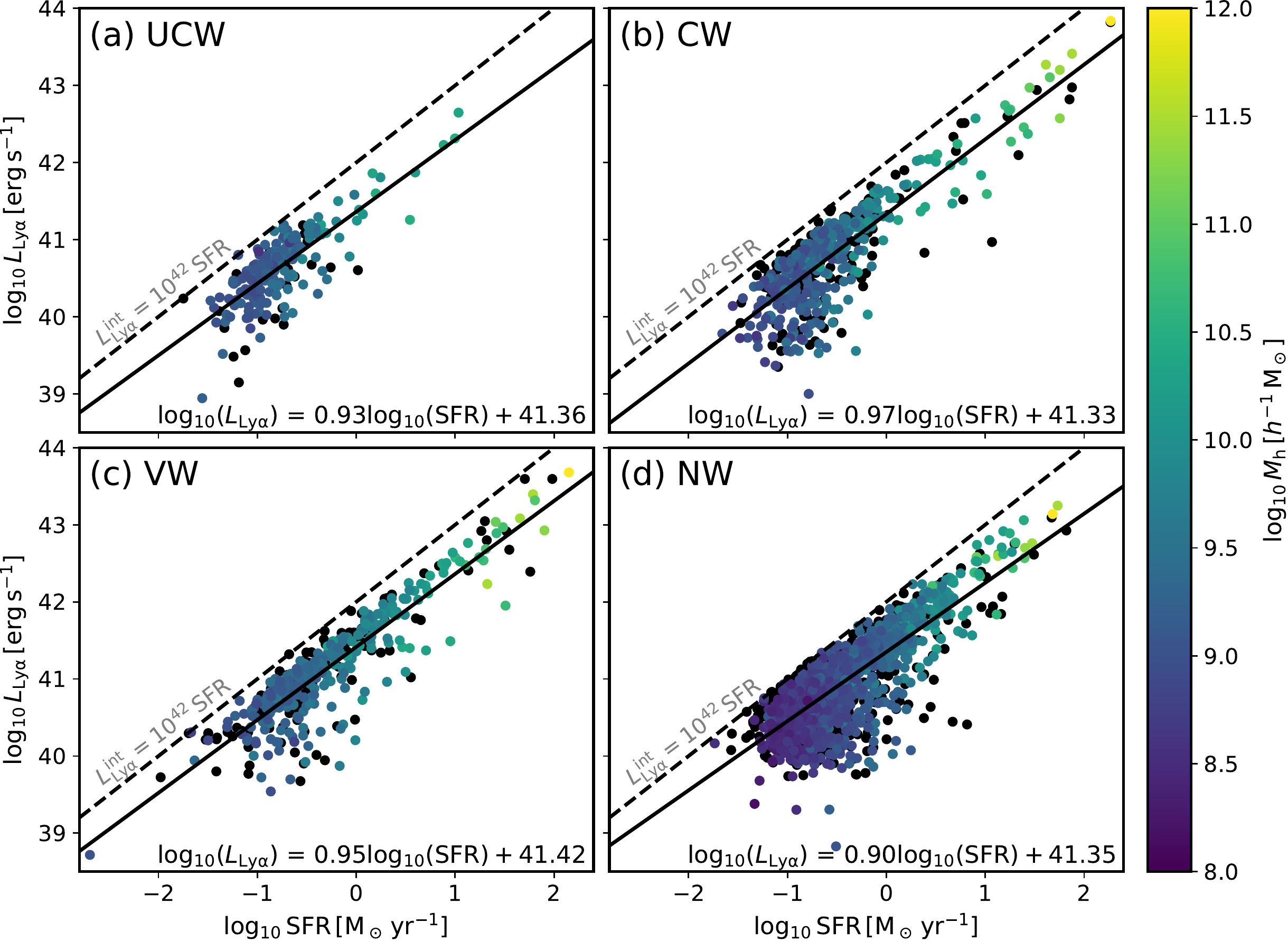}
\caption{Relation between apparent \lya luminosity and star formation rate for galaxies in the different simulation runs at $z=6.6$. Each symbol represents an individual galaxy. The color of the symbol indicates the mass of the DM halo hosting the galaxy if it is the brightest among galaxies residing in that halo. Other galaxies residing in the same halo are shown as black symbols. The solid lines correspond to the linear regression fits whose coefficients are indicated in the bottom right of each panel. The dashed lines show the relation between \emph{intrinsic} \lya luminosity, $L_{\rm Ly\alpha}^{\rm int}$, and SFR that we assume in our radiative transfer model (Eq. \ref{eq:lya_lum}).}
\label{fig:L-sfr}
\end{figure*}

\begin{figure*}
\centering
\includegraphics[width=0.9\linewidth]{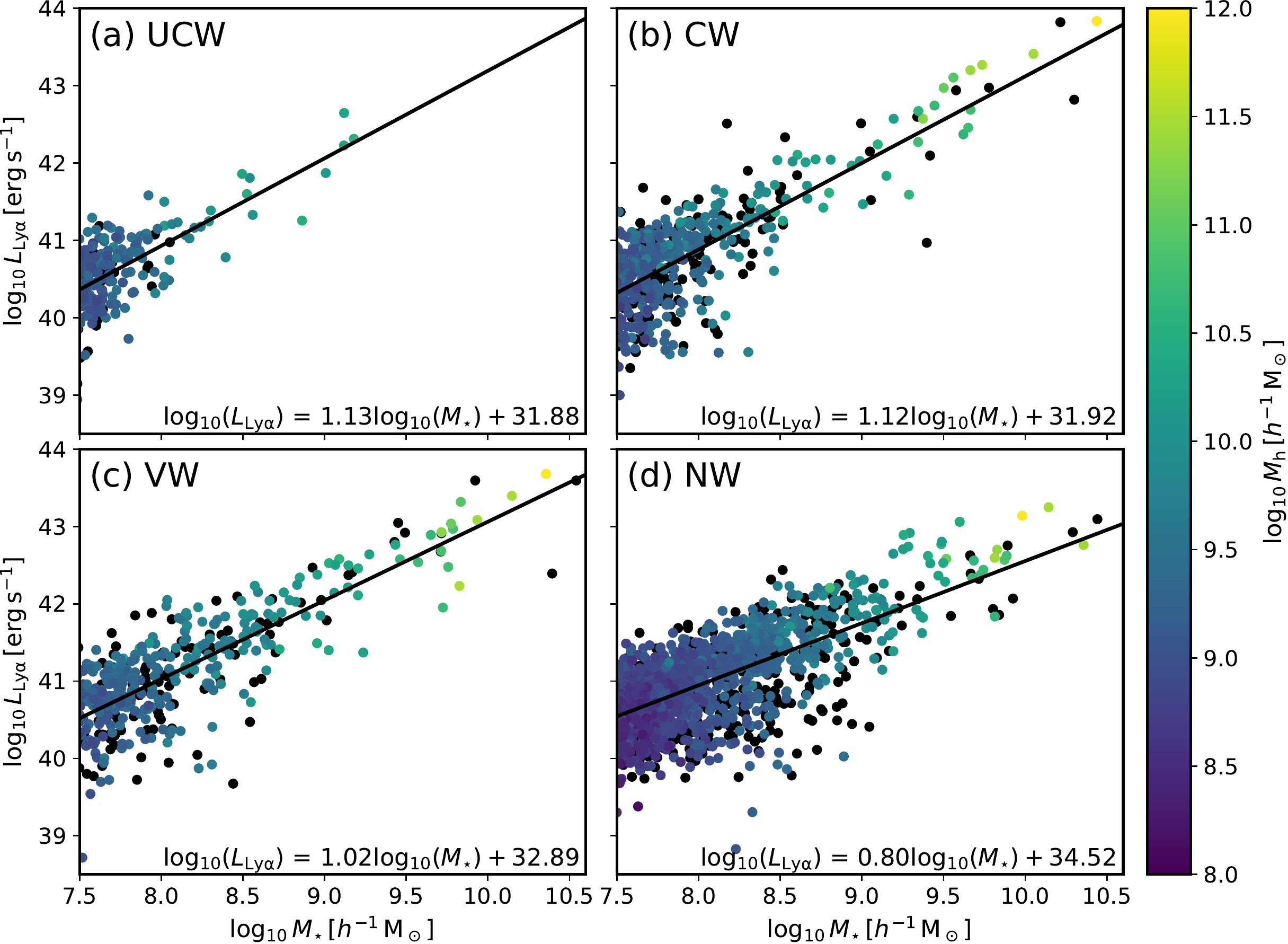}
\caption{Same as Fig. \ref{fig:L-sfr} except that the apparent \lya luminosity 
is now shown as a function of the galaxy stellar mass, $M_{\star}$. The lower limit on the $x$-axis is set to $M_{\star} \sim 3\,\times 10^7\,h^{-1}\,\mathrm{M}_\odot$, corresponding to the mass above which galaxies are properly resolved in our simulations.}
\label{fig:L-Mstar}
\end{figure*}

The dependence of the \lya emission from high-redshift LAEs on galaxy properties is poorly understood for $z\gtrsim 6$, even more so for objects residing in extremely biased regions such as the overdense region in our CR simulations. Here, we explore the relation between apparent \lya luminosity and two key galaxy properties: stellar mass and SFR. We present the predictions of our models as a function of environment (by comparing UCW and CW simulations) and as a function of wind models (by comparing CW, VW, and NW).

In order to estimate the apparent \lya luminosity $L_{\rm Ly\alpha}$ of a galaxy in our simulation, we use the output of the radiative transfer calculations that provides the full spatial and spectral distribution of the \lya emission (Section \ref{sec:methods}). However, as already mentioned, galaxies do not appear as point sources but have an extended diffuse emission around them as a result of the resonant scattering of \lya photons. To include this diffuse component, we select all pixels in the \lya image that fall within a circular region of radius 1$\arcsec$ around the galaxy, which is the typical aperture size for narrowband imaging surveys of LAEs at high redshifts \citep[e.g.,][]{Konno2018}. The apparent \lya luminosity of the galaxy is calculated as the total \lya emission in this region.

It is important to note that the \lya radiative transfer calculations stop following the scattering of \lya photons after they have escaped the simulation volume. Since, in our case, this volume is very small, the photons on the blue side of the line do not travel far enough to be able to redshift into resonance. For this reason, the majority of the \lya spectra predicted by our radiative transfer model have a blue component. However, we only consider the red part of the \lya spectra when computing apparent fluxes as the blue component should be heavily suppressed once photons reach the neutral IGM on scales larger that our simulation box. 

\begin{figure*}
\centering
\includegraphics[width=0.9\linewidth]{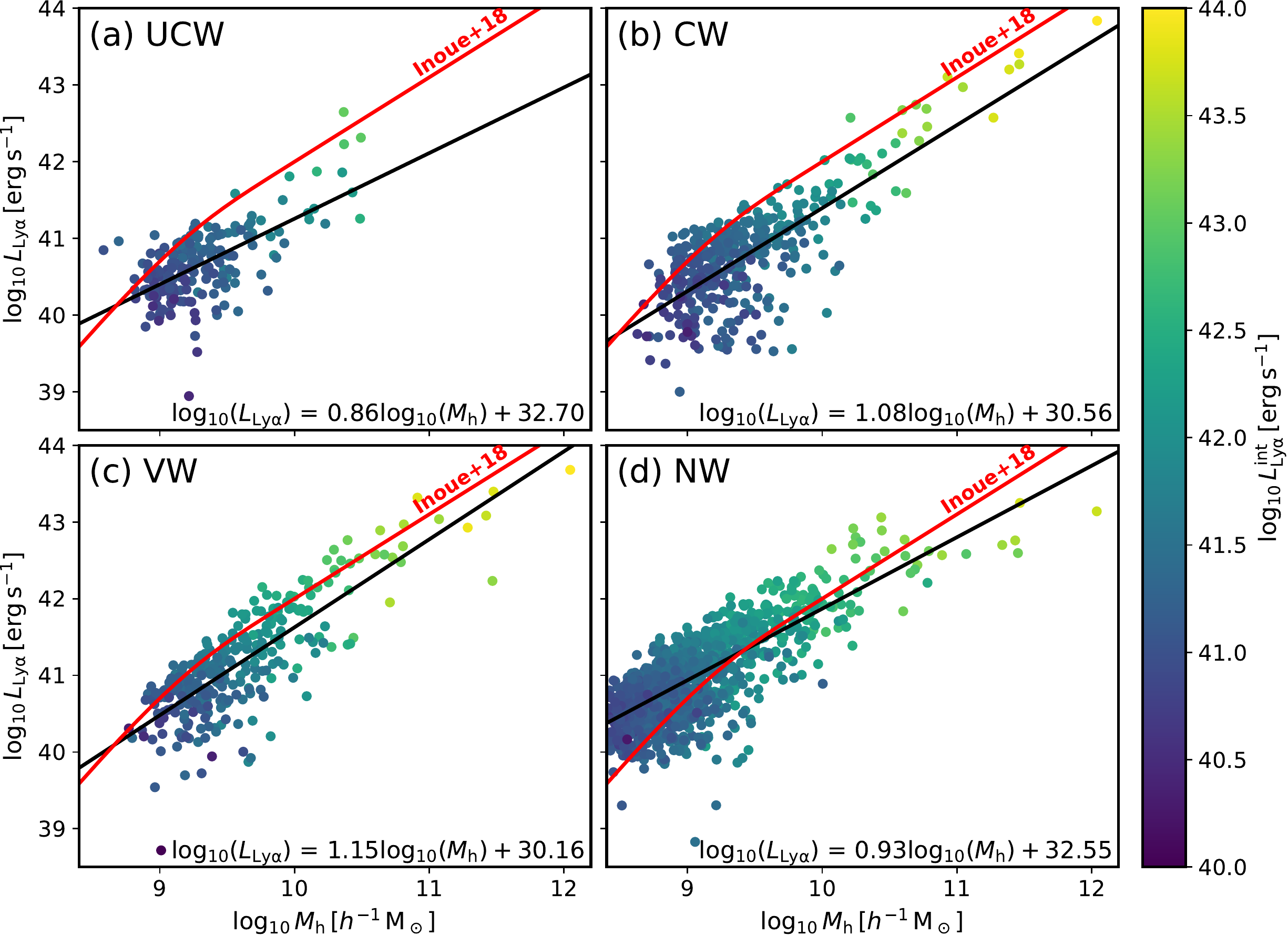}
\caption{Relation between apparent \lya luminosity and halo mass for \emph{central} galaxies in our simulation runs. The color of the symbols indicate the intrinsic \lya luminosity of the galaxy. The black solid lines correspond to the best linear fit in $\log_{10}(L_{\rm Ly\alpha})-\log_{10}(M_{\rm h})$, with $L_{\rm Ly\alpha}$ in $\mathrm{erg}\,\mathrm{s}^{-1}$ and $M_{\rm h}$ in $h^{-1}\,\mathrm{M}_\odot$. The red solid line is the mean $L_{\rm Ly\alpha}$-$M_{\rm h}$ relation from the LAE model of \citet{Inoue2018}. The differences in the $L_{\rm Ly\alpha}$-$M_{\rm h}$ relation between the CW, VW and NW models are driven by the effects of galactic winds on apparent \lya luminosity. }
\label{fig:L-Mh}
\end{figure*}

Fig. \ref{fig:L-sfr} shows the relation between $L_{\rm Ly\alpha}$ and SFR. We have plotted the prediction for each model in its own panel as indicated by the labels on their top left corners. Each filled circle represents an individual \lya source (i.e. galaxy). The color of the symbol indicates the mass, $M_{\rm h}$, of the DM halo hosting the galaxy. In our simulations, multiple galaxies can be identified as member of the same DM halo. In order to properly assign a single galaxy per halo, we define the central galaxy of a halo to be the brightest object among galaxies residing in that halo. Other galaxies in the same halo are shown as black symbols on Fig. \ref{fig:L-sfr}. This definition allows us to study the \lya luminosity-halo mass relation for central galaxies (in Section \ref{sec:lya-halo}) without introducing contamination from satellite galaxies. The solid lines correspond to the best linear fit in $\log_{10}(L_{\rm Ly\alpha})-\log_{10}(\mathrm{SFR})$, with $L_{\rm Ly\alpha}$ in $\mathrm{erg}\,\mathrm{s}^{-1}$ and SFR in $\mathrm{M}_\odot\,\mathrm{yr}^{-1}$. The coefficients of the regression are shown in the bottom right corner of each panel. 

Overall, we find that $L_{\rm Ly\alpha}$ increases with SFR in all models but with a shallower slope than the relation assumed between \emph{intrinsic} \lya luminosity and SFR, shown as a dashed line in each panel (Eq. \ref{eq:lya_lum}). The relation also exhibits some scatter in $L_{\rm Ly\alpha}$ at fixed SFR in all our models. In the CR simulations, the rms scatter in $\log_{10}(L_{\rm Ly\alpha})$ is $\sim 0.3-0.4\,\mathrm{dex}$ for galaxies with SFR in the range $0.1-10\,\mathrm{M}_\odot\,\mathrm{yr}^{-1}$. In the UCW model, most galaxies have SFRs below $10\,\mathrm{M}_\odot\,\mathrm{yr}^{-1}$ but the scatter of the relation for these objects remains roughly the same as in the CR models.

The brightest objects in the CW and VW models reach $L_{\rm Ly\alpha} \gtorder 10^{43.5}\,\mathrm{erg}\,\mathrm{s}^{-1}$, consistent with the recent luminous \lya emitters observed at $z\sim 6.6$ with HST \citep{Hu2016,Songaila2018} and with the bright end of the \lya luminosity function (LF) of LAEs sampled by Subaru/HSC \citep{Matthee2015,Konno2018}. In our simulations, these luminous galaxies have SFR between $50-200\,\mathrm{M}_\odot\,\mathrm{yr}^{-1}$ and correspond to objects that form within the central, massive DM halo. The brightest galaxy is found in the CW model with an apparent \lya luminosity of $L_{\rm Ly\alpha} \sim 10^{43.85} {\rm erg\, s^{-1}}$ and corresponding $\mathrm{SFR} \sim 185 \,\mathrm{M}_\odot\,\mathrm{yr}^{-1}$. According to Eq. \ref{eq:lya_lum}, the intrinsic \lya luminosity of this source is $L_{\rm Ly\alpha}^{\rm int} \sim 10^{44.27}\,\mathrm{erg}\,\mathrm{s}^{-1}$. Comparing the intrinsic and apparent \lya luminosities, the fraction of transmitted flux after being processed by the IGM amounts to roughly $40\%$ for this galaxy. The higher \lya luminosity found in the CW model for the central galaxy results mainly from two effects: (i) the same galaxy in the VW and NW runs have lower SFRs ($95$ and $47\,\mathrm{M}_\odot\,\mathrm{yr}^{-1}$ respectively) and (ii) the ionized region around the galaxy is larger in CW than in the other two models (as seen on Fig. \ref{fig:slice_maps}).

The relations between apparent \lya luminosity and stellar mass are shown on Fig. \ref{fig:L-Mstar}, with a similar presentation as in Fig. \ref{fig:L-sfr}. The solid lines now indicate the best linear fit in $\log_{10}(L_{\rm Ly\alpha})-\log_{10}(M_\star)$ with $M_\star$ in $h^{-1}\,\mathrm{M}_\odot$.
We find again some scatter in apparent \lya luminosity at fixed $M_\star$. For the CR simulations, the scatter in $\log_{10}(L_{\rm Ly\alpha})$ for galaxies in the mass range $10^{7.5}-10^{9.5}\,h^{-1}\,\mathrm{M}_\odot\,$ is slightly larger ($\sim 0.4-0.5\,\mathrm{dex}$) than the scatter of the $L_{\rm Ly\alpha}-\mathrm{SFR}$ relation.  The linear fits on \ref{fig:L-Mstar} show that $L_{\rm Ly\alpha}$ increases close to linearly with $M_\star$ in the CW and VW models. The relation has a shallower slope in the NW model which is likely the result of massive galaxies having smaller SFRs (at fixed $M_\star$) in NW compared to CW and VW at this redshift. Interestingly, the relation for the UCW run is very similar to that of CW and VW which indicates that the large-scale environment has little effect on the apparent \lya luminosity of most galaxies in our simulations \emph{at a given stellar mass}\footnote{Note that this does not contradict the results presented in Section \ref{sec:maingal} because (i) the relations shown here are primarily driven by the large number of low and intermediate-mass galaxies while we studied specifically the most massive galaxy in Section \ref{sec:maingal}, and (ii) the two galaxies compared in Section \ref{sec:maingal} have similar host \emph{halo masses} but different \emph{stellar masses} as a result of different formation histories in the CW and UCW simulations.}.

%%%%%%%%%%%%%%%%%%%%%%%%%%%%%%%%%%%%%%%%%%%%%%%%%%
\subsection{\lya luminosity-halo mass relation}
\label{sec:lya-halo}

\begin{figure*}
\centering
\includegraphics[width=0.85\linewidth]{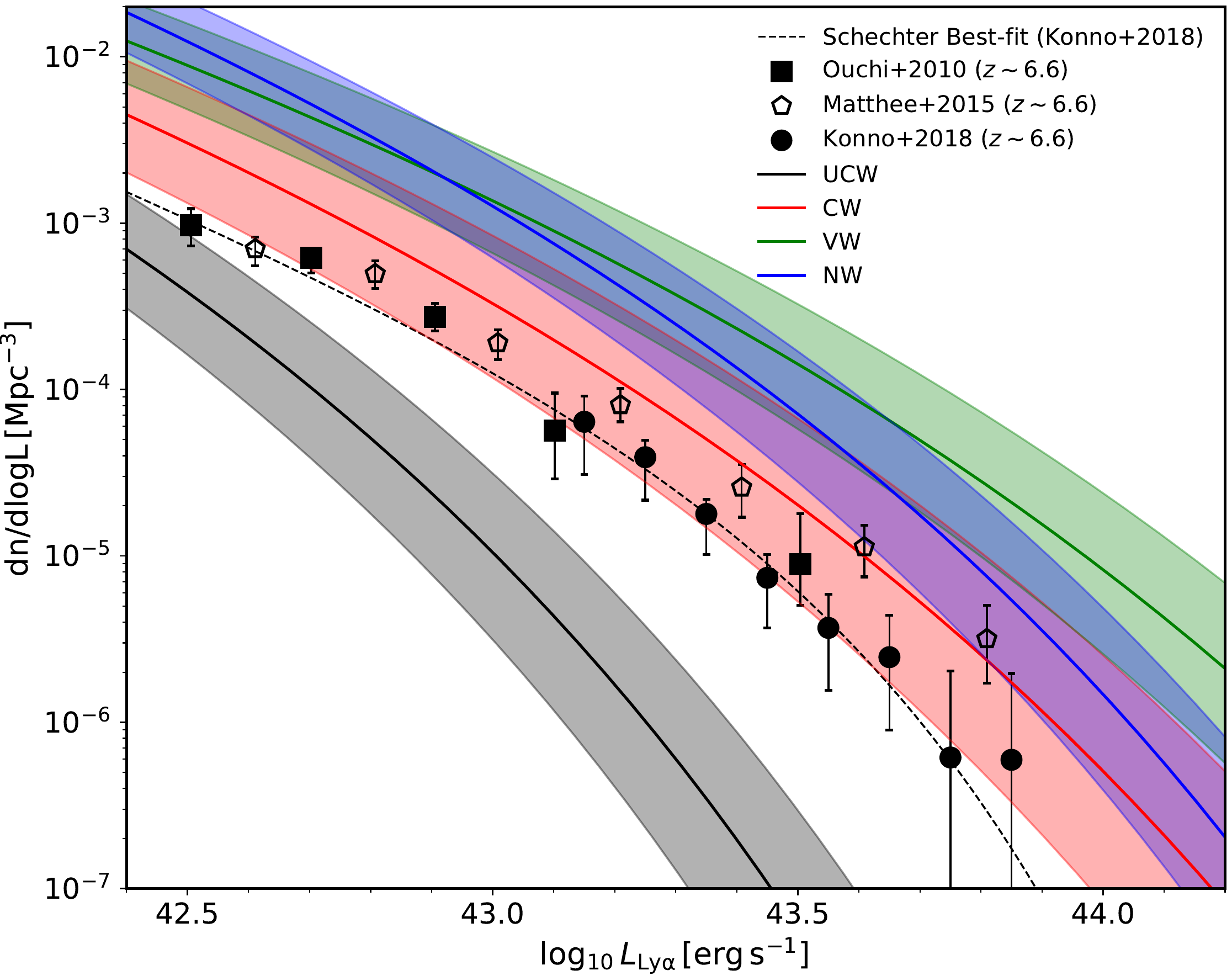}
\caption{Comparison between the \lya luminosity functions (LFs) calculated using the $L_{\rm Ly\alpha}$-$M_{\rm h}$ relation predicted by our radiative transfer calculations in the unconstrained (UCW) and constrained (CW,VW, and NW) simulations runs with observational data. The \lya LFs are obtained by convolving the $L_{\rm Ly\alpha}$-$M_{\rm h}$ relation measured in the simulations with the theoretical halo mass function (MF) from \citet{Sheth2002}. The shaded region indicates the $1\sigma$ uncertainty due to the average scatter in the $L_{\rm Ly\alpha}$-$M_{\rm h}$ relation. The data points show recent estimates of the \lya LF at $z$=6.6 from \citet{Ouchi2010}, \citet{Matthee2015}, and \citet{Konno2018}.}
\label{fig:LyaLF}
\end{figure*}

In Fig. \ref{fig:L-sfr} and Fig. \ref{fig:L-Mstar}, it is hinted that the relation between $L_{\rm Ly\alpha}$ and $M_{\rm h}$ is approximately monotonic, with brighter object residing in more massive haloes. Here, we explore in more details the connection between halo mass and apparent \lya luminosity for galaxies in the different simulation runs. This can be useful not only to gain insight on the type of haloes that are hosting high-$z$ LAEs but also for calibrating semi-analytical models or for producing mock LAEs samples from N-body simulations. 

The relation between $L_{\rm Ly\alpha}$ and $M_{\rm h}$ is shown in Fig. \ref{fig:L-Mh}. We only show the relation for central galaxies (identified as the brightest galaxy in a given halo, as already mentioned) in order to avoid contamination from satellites. The main reason for not including satellite galaxies is that we do not identify DM substructures as separate DM haloes in the simulations. This means that satellite galaxies are assigned to the parent DM halo in which they orbit instead of their own DM (sub)-halo, which results in the host halo mass being overestimated for these objects. By not removing these galaxies from our analysis, we would therefore introduce some bias in the $L_{\rm Ly\alpha}$-$M_{\rm h}$ relation, especially for the CR simulations where more than $40$ galaxies are found within the central, massive halo.

Fig. \ref{fig:L-Mh} shows that the $L_{\rm Ly\alpha}$-$M_{\rm h}$ relation is similar between the CW and VW models, while it is shallower in the NW model. Given that the CR simulations only differ by the type of wind model implemented, the formation and evolution of DM haloes is the same between the CW, VW and NW models (see the DM halo mass function shown in Fig. 3 in S16). On the other hand, we showed in S16 that, at $z\sim 6-7$, galaxies  of a given stellar mass in the NW model have, on average, lower SFRs than in the other two wind models, which means that these galaxies are intrinsically fainter compared to those in the CW and VW models. In principle, this could explain the differences we find in the $L_{\rm Ly\alpha}$-$M_{\rm h}$ relation. However, if the reduction in the SFR of galaxies in the NW model compared to CW and VW cases is the same for every galaxy, then that would not affect the slope of the relation, only its normalization. In fact, we find that the relation between intrinsic \lya luminosity (indicated by the color of the symbols in Fig. \ref{fig:L-Mh}) and halo mass appears to be consistent between VW and NW models. The differences in the $L_{\rm Ly\alpha}$-$M_{\rm h}$ relation between those models are therefore entirely driven by the presence of galactic winds and their effects on apparent \lya luminosity in the CW and VW models.

The black solid lines in Fig. \ref{fig:L-Mh} indicate the best linear fit in $\log_{10}(L_{\rm Ly\alpha})-\log_{10}(M_{\rm h})$, with $L_{\rm Ly\alpha}$ in $\mathrm{erg}\,\mathrm{s}^{-1}$ and $M_{\rm h}$ in $h^{-1}\,\mathrm{M}_\odot$. These fitting formulas can be used to estimate the apparent \lya luminosity of galaxies as a function of halo mass. This is valuable for semi-analytical models and for quickly generating mock LAE samples from cosmological simulations to compare with observational data at $z \gtorder 6$. In the next section, we demonstrate how we can apply the formulas to calculate mock \lya luminosity functions from our simulations. These fits also show that the $L_{\rm Ly\alpha}$-$M_{\rm h}$ relation differs between UCW and CW models, which implies that, \emph{at fixed halo mass}, the apparent \lya luminosity of a galaxy is function of the large-scale environment. This is consistent with the comparison presented in Section \ref{sec:maingal} between the most massive galaxy in the UCW model and its counterpart (residing in the same halo) in the CW model.

We also show in Fig. \ref{fig:L-Mh} the mean $L_{\rm Ly\alpha}$-$M_{\rm h}$ relation used by \citet{Inoue2018} to model LAEs in a large cosmological reionization simulation (red solid lines). Note that we show here the relation corresponding to their Eq. 6 which relates halo mass to \emph{intrinsic} \lya luminosity. To calculate the apparent \lya luminosity of a galaxy in their simulation, they have to account for the line transfer through the ISM (described by an effective \lya escape fraction in their model) as well as for the transmission through the IGM which is obtained by computing the \lya optical depth along a few directions. In our case, the \lya transmission through the CGM and IGM is obtained directly from the radiative transfer calculations. The slope of the relation used by \citet{Inoue2018} is consistent with our result for haloes with $M_{\rm h}>10^9\,h^{-1}\,\mathrm{M}_\odot$ in the CW and VW models. In particular, the relation we obtain for the VW model is very similar to theirs, while the one we find in the NW model deviates significantly from it. This is a good indication that galactic winds are a necessary ingredient to include in order to properly model the $L_{\rm Ly\alpha}$-$M_{\rm h}$ relation at $z\gtorder 6$.

The most \lya luminous (and most massive) galaxy in the CW and VW models is hosted by the central, massive DM halo with a mass of $M_{\rm h}\sim\,10^{12}\,h^{-1}\,\mathrm{M}_\odot$. In the NW model, the same galaxy is only the third brightest object while the most luminous galaxy in this model, having $L_{\rm Ly\alpha}\sim \,10^{43.2}\,{\rm erg\,s^{-1}}$, is found within the second most massive halo with a mass of $M_{\rm h}\sim\,3\times\,10^{11}\,h^{-1}\,\mathrm{M}_\odot$. As mentioned in the previous section, the apparent \lya luminosity of the most massive galaxy in the CW and VW models matches well that of observed luminous LAEs at $z \sim 6.6$. We stress that this is a natural outcome of our \lya modeling as we have not calibrated our radiative transfer calculations against observational data. Our results thus suggest that the host haloes of observed luminous LAEs at $z\sim 6.6$ have masses $M_{\rm h}\sim\,10^{12}\,h^{-1}\,\mathrm{M}_\odot$ and are found within rare, highly overdense, ionized regions reminiscent of the one that forms in the CR simulations.

%%%%%%%%%%%%%%%%%%%%%%%%%%%%%%%%%%%%%%%%%%%%%%%%%%%%%%%%%%%%%%%%%%%
\section{Discussion}
\label{sec:discussion}
%%%%%%%%%%%%%%%%%%%%%%%%%%%%%%%%%%%%%%%%%%%%%%%%%%%%%%%%%%%%%%%%%%%

%%%%%%%%%%%%%%%%%%%%%%%%%%%%%%%%%%%%%%%%%%%%%%%%%%
\subsection{\lya luminosity functions at $z \sim 6.6$}
\label{sec:lyaLF}

A key observable quantity in the study of LAEs during the reionization epoch is the \lya luminosity function (LF), which provides the observed number density of LAEs as a function of their apparent \lya luminosity. Recent large narrowband LAE surveys have shown that, while the faint-end of the \lya LF evolves rapidly between $z=6.6$ and $z=7.3$, the number density of luminous LAEs ($\gtorder 10^{43.5}\,{\rm erg\,s^{-1}}$) appears to stay relatively constant in the same redshift interval \citep{Ouchi2010,Ono2012,Matthee2015,Zheng2017,Konno2018}. In addition, the observed bright end of the \lya LF at $z\sim 6.6-7$ appears in excess of the best-fit Schechter function obtained from fitting lower luminosity LAEs. This has been interpreted as a signature of patchy reionization, implying that bright galaxies are found preferentially in ionized, overdense regions which more easily transmit the \lya line \citep{Zheng2017,Weinberger2018}.

To compare these observations with results from our \lya radiative transfer calculations, we want to estimate the \lya LF in our simulations. We could, in principle, compute the LFs directly from the simulations since our radiative transfer calculations provide the apparent \lya luminosity of each galaxy. However, there are two problems with this approach. First, we only have a handful of luminous galaxies in our simulations which means than the bright end of our LFs will be dominated by Poisson noise and will not be properly resolved. Second, the volume covered by our simulations is much smaller than the typical size of narrowband imaging surveys at high $z$. For comparison, the recent SILVERRUSH program \citep{Ouchi2017} from HSC/Subaru covered $\geq 15\, \mathrm{deg}^2$ on the sky, corresponding to $0.3-0.5\, \mathrm{Gpc}^2$ at $z\sim 6-7$, while our simulations effectively only cover $\sim 40\,\mathrm{Mpc}^2$.

For these reasons, we decide to take a different approach and use the $L_{\rm Ly\alpha}$-$M_{\rm h}$ relations that we obtain from our radiative transfer calculations (Section \ref{sec:lya-halo}). To compute the \lya LF for one of our simulation run, we convolve the $L_{\rm Ly\alpha}$-$M_{\rm h}$ relation with the theoretical halo mass function (MF) from \citet{Sheth2002}. We use the best-fit relation shown in Fig. \ref{fig:L-Mh} to calculate the apparent \lya luminosity of LAEs as a function of halo mass and use the theoretical halo MF to get the expected number density of halos of a given mass. We also account for the dispersion in the $L_{\rm Ly\alpha}$-$M_{\rm h}$ relation by measuring the scatter around the best-fit values as a function of halo mass and averaging the result over all masses to estimate the mean uncertainty associated with the relation. 

We obtain an average scatter of $0.13\,\mathrm{dex}$, $0.20\,\mathrm{dex}$, $0.17\,\mathrm{dex}$ and $0.14\,\mathrm{dex}$ for the UCW, CW, VW and NW models respectively. Note that the $L_{\rm Ly\alpha}$-$M_{\rm h}$ relations calculated in Section \ref{sec:lya-halo} refer only to the central galaxies. As a result, the \lya LFs obtained using these relations do not include contribution from satellites. We expect this contribution to be small and affect mostly
the faint-end of the LF \citep[e.g.,][]{vale04,vale06}, which is mostly unresolved by our simulations.

The \lya LFs calculated with this method are shown in Fig. \ref{fig:LyaLF}. The solid lines indicate the LF obtained from using the fitting formulas given in Section \ref{sec:lya-halo}. The shaded regions correspond to the $1\sigma$ uncertainty in the LFs coming from the uncertainty in the $L_{\rm Ly\alpha}$-$M_{\rm h}$ relation (the average scatter discussed above). The data points are recent estimate of the \lya LF at $z=6$ from \citet{Ouchi2010}, \citet{Matthee2015}, and \citet{Konno2018}. The best Schechter fit to the observed \lya LF in the range $\log_{10}\,L_{\rm Ly\alpha}\sim\,42.4-44\,{\rm erg\, s^{-1}}$ from \citet{Konno2018} is shown as a dashed line in Fig \ref{fig:LyaLF}. We find that the LF obtained in the CW model is remarkably consistent with the observed one, both at the faint-end and at the bright-end. On the other hand, the LF for UCW severely underestimates the number density of bright LAEs by several orders of magnitude. This is most likely due to the lack of luminous objects (i.e massive haloes) in the UCW model which makes the $L_{\rm Ly\alpha}$-$M_{\rm h}$ relation calculated in the simulation unreliable for computing the bright end of the LF. 

The \lya LFs in the CW and VW models reproduce well the shape of the best Schechter fit, both the faint-end slope and the exponential cutoff at the bright-end, while the LF in the NW model appears to be steeper than the Schechter fit. This means that our radiative transfer calculations were able to properly capture the correct scaling between $L_{\rm Ly\alpha}$ and $M_{\rm h}$ in the CW and VW models. However, the LFs in the VW and NW models deviate from the observed one over the entire luminosity range of the data. We speculate that this may be because we do not account for dust in our radiative transfer calculations. While we argue in Appendix \ref{app:dust} that dust is probably not important for the \lya line transfer through the IGM in our simulations, it is likely to be an important factor in reducing the escape fraction, $f_{\rm esc}^{\rm Ly\alpha}$, of \lya photons out of galaxies, especially massive ones. This is also supported by the fact that galaxies in the CW model are the most efficient at ejecting metals from their ISM into their surrounding IGM (as discussed in S16), which means that these galaxies should be the least affected by our omission of the dust. This simple explanation is consistent with the fact that the LF in the CW model has the closest match with the observed LF. 

A quick way to assess the effect of dust on ISM scales in our radiative transfer calculations is by estimating the value of the \lya escape fraction, $f_{\rm esc}^{\rm Ly\alpha}$, required to match the \lya LFs of our models with the observed one. For simplicity, here we assume that $f_{\rm esc}^{\rm Ly\alpha}$ is independent
of halo mass. Comparing the \lya LFs obtained in our models with the observed LF shown in Fig. \ref{fig:LyaLF}, we find that $f_{\rm esc}^{\rm Ly\alpha}\sim0.50,0.25$ and $0.39$ for the CW, VW and NW models respectively. These values are broadly consistent with the ones measured in detailed \lya radiative transfer calculations on sub-ISM scales with dust \citep[see, e.g., the recent study by][]{Smith2018}.

\subsection{Comparison with an observed galaxy overdensity at $z=7$}
\label{subsec:observ_overdens}

Recent deep spectroscopic follow-up observations of colour-selected galaxies in the Bremer Deep Field (BDF) have revealed LAEs associated with a galaxy overdensity at $z\sim 7$ \citep[][hereafter C18]{Castellano2016,Castellano2018}. These observations are ideal to compare the results of our CR simulations with real data. The galaxy sample presented in C18 corresponds to 16 colour-selected galaxy candidates previously observed with HST and believed to be associated with a galaxy overdensity at $z\sim 7$. Out of these 16 objects, only 3 of them are identified as bright LAEs with \lya equivalent width (EW) $\geq 50$ \AA. The other objects have faint \lya emission below the detection limit, with \lya EW $\leq 25$ \AA. The newly confirmed LAE from C18 has a broad, asymmetric \lya line profile with a line width of $\mathrm{FWHM}=240\,{\rm km\, s^{-1}}$ and a line flux of $1.85\times 10^{-17}\,{\rm erg\, s^{-1}\, cm^{-2}}$. In our CR simulations, the brightest galaxy has a \lya line flux (redward of systemic redshift) of $1.31\times10^{-16}$, $7.51\times10^{-17}$ and $2.37\times10^{-17}\,{\rm erg\,s^{-1}\,cm^{-2}}$ for CW, VW and NW respectively. As already mentioned, the most massive galaxy in the UCW simulation has $\sim 20$ times lower SFR than the one in the CW model and is thus much fainter with a line flux of $3.93\times10^{-18}\,{\rm erg\,s^{-1}\,cm^{-2}}$.

\begin{figure}
\includegraphics[width=\linewidth]{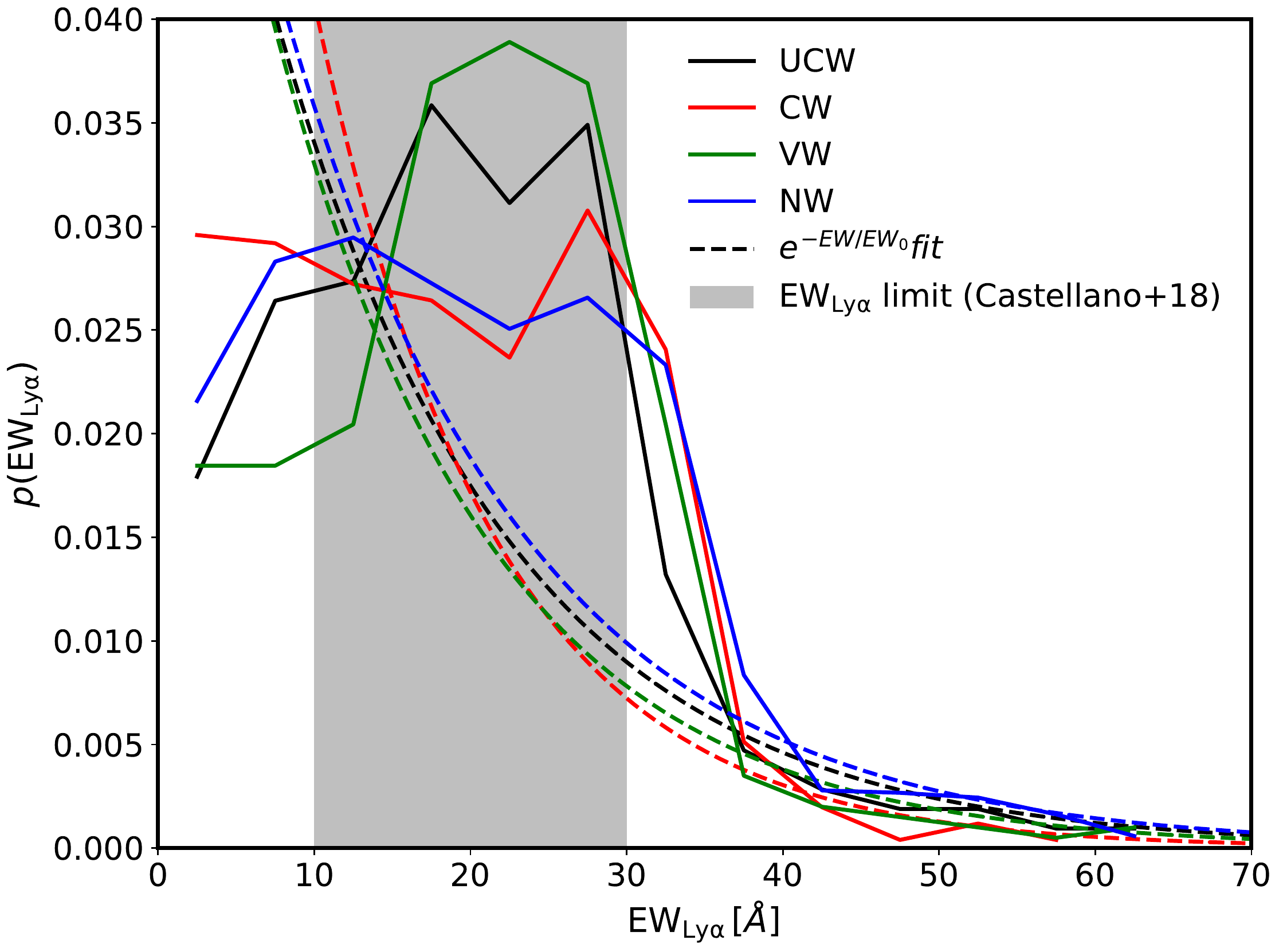}
\caption{Distributions of lya EWs in the CW (red), VW (green), NW (blue) and UCW (black) simulation runs. The solid lines are the distribution of \lya EWs obtained from the simulations. The dashed lines correspond to the exponential fit of the form $p(\mathrm{EW}_{\rm Ly\alpha})\propto e^{-\mathrm{EW}_{\rm Ly\alpha}/\mathrm{EW}_0}$ to the distributions. The gray area indicates the detection limits in $\mathrm{EW}_{\rm Ly\alpha}$ at $z\sim 6-7$ from \citet{Castellano2018}.
}
\label{fig:lya_EW}
\end{figure}

The most intriguing aspect of the C18 observations is the lack of apparent \lya emission for continuum-faint galaxies located in an overdense environment. C18 have shown that the detection of the three bright LAEs is in agreement with the galaxy overdensity being associated with an ionized region, which enhances the visibility of the \lya line. On the other hand, the lack of detected \lya emission in fainter galaxies contradicts this simple picture. We have tested the \lya visibility of our simulated galaxies by computing their \lya EWs. 

For each galaxy, the \lya EW is calculated as 
\begin{equation}
\mathrm{EW}_{\rm Ly\alpha} \equiv \frac{L_{\rm Ly\alpha}}{L_{\lambda}^{\rm cont}},
\label{eq:lya_ew}
\end{equation}
where $L_{\rm Ly\alpha}$ is the apparent \lya luminosity obtained from our model and $L_{\lambda}^{\rm cont}$ is the continuum luminosity per unit wavelength at the \lya frequency. In order to estimate the continuum, we first compute the specific UV luminosity at $1500$ \AA, $L_{\nu}^{\rm UV}$, which can be derived from the galaxy SFR assuming Salpeter IMF and solar metallicity\citep{Madau1998} as
\begin{equation}
L_{\nu}^{\rm UV} = 8\times10^{27} \left(\frac{\mathrm{SFR}}{\mathrm{M}_\odot\,\mathrm{yr}^{-1}}\right){\rm\,erg\,s^{-1}\,Hz^{-1}}.
\label{eq:L_UV_nu}
\end{equation}
We assume that the galaxy spectrum in the UV is of the form $L_{\lambda}\propto \lambda^{\beta}$ and take $\beta = -2$ for our calculations.
Under this assumption, the continuum luminosity at the \lya frequency can be expressed as

\begin{equation}
L_{\lambda}^{\rm cont} = 1.62\times\,10^{40}\,\left(\frac{\mathrm{SFR}}{\mathrm{M}_\odot\,\mathrm{yr}^{-1}}\right){\rm\, erg\,s^{-1}\,}\textrm{\AA}^{-1},
\label{eq:L_UV_lambda}
\end{equation}
and the \lya EW can be calculated as a function of SFR from 
\begin{equation}
\mathrm{EW}_{\rm Ly\alpha} = 61.62\left(\frac{\mathrm{SFR}}{\mathrm{M}_\odot\,\mathrm{yr}^{-1}}\right)^{-1}\left(\frac{L_{\rm Ly\alpha}}{10^{42}\,\rm erg\,s^{-1}}\right)\textrm{\AA}.
\label{eq:EW-SFR}
\end{equation}

The distributions of \lya EW in the different simulation runs are shown in Fig. \ref{fig:lya_EW}. The solid lines correspond to the EW distributions calculated according to Eq. \ref{eq:EW-SFR} for galaxies in the simulations. The dashed lines correspond to an exponential fit of the form $p(\mathrm{EW}_{\rm Ly\alpha})\propto e^{-\mathrm{EW}_{\rm Ly\alpha}/\mathrm{EW}_0}$ to the distributions, where $\mathrm{EW}_0$ is the scale length of the distribution. The gray area indicates the detection limits in $\mathrm{EW}_{\rm Ly\alpha}$ at $z\sim 6-7$ from C18. The largest \lya EWs found in our simulations have values $\sim 61$\AA\, $56$\AA\, $62$\AA\ and $61$\AA\ for the UCW, CW, VW and NW models respectively. All the simulation runs show very similar EW distributions with scale lengths of $\sim 15$\AA\, $12$\AA\, $ 14$\AA\ and $16$\AA.

Interestingly, we find that the majority of our simulated galaxies have \lya EWs below $\sim 30$\AA, corresponding to the detection limits in the C18 observations. In the case of the CR simulations, a small number of objects could be identified as strong LAEs based on their \lya EWs in the C18 observations ($\mathrm{EW}_{\rm Ly\alpha}\gtorder 50$\AA), although galaxies in the UCW simulation also have EWs above this limit. Based on these results, we tentatively suggest that the distribution of \lya EWs does not depend strongly on the large-scale environment at $z\sim 6.6$. This could explain why the C18 observations have yielded only a handful of luminous LAEs despite targeting an overdensity of colour-selected galaxy candidates.

\subsection{Comparison with observations of ultra-luminous LAEs at $z=6.6$}

\begin{figure*}
\includegraphics[width=\linewidth]{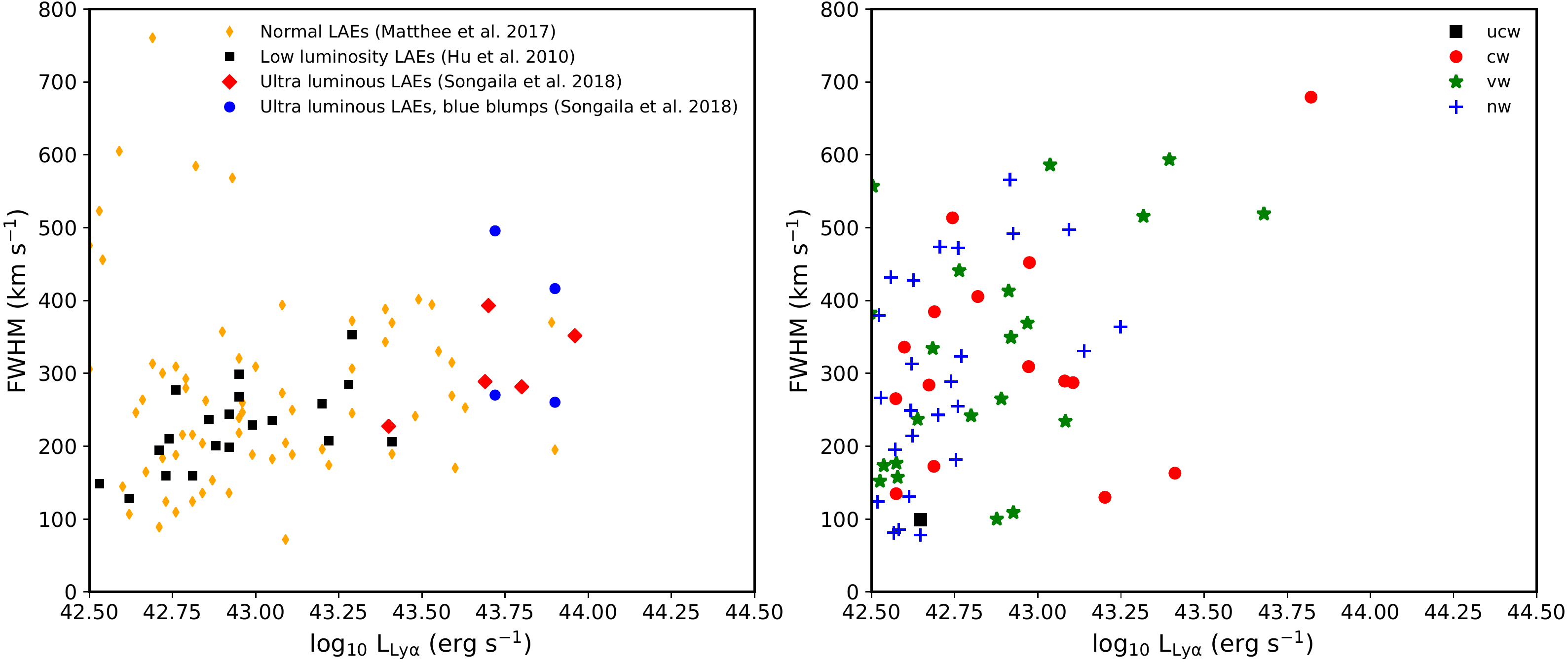}
\caption{Width of the \lya line as a function of apparent \lya luminosity for observed normal and ultra-luminous LAEs (left) and for our simulated galaxies (right). The left panel is showing the recent ultra luminous LAEs detected by \citet{Songaila2018} (including the COLA1 galaxy) with a similar presentation as their Fig. 13. Average and low-luminosity LAE samples from \citet{Hu2010} and \citet{Matthee2017} are also shown, as in \citet{Songaila2018}. The right panel shows the result for our simulated galaxies in the different simulations runs.}
\label{fig:lya_fwhm}
\end{figure*}

As mentioned previously, the most massive galaxies in the CR simulations are very bright LAEs. This is especially true in the CW and VW models where the most massive galaxy has a line flux close to $\sim 10^{-16}\,{\rm erg\, s^{-1}\, cm^{-2}}$ in the red peak and a corresponding apparent \lya luminosity of $L_{\rm Ly\alpha} = 10^{43.84}$ (CW) and $10^{43.59}$ (VW) ${\rm erg\, s^{-1}}$. These values are in the range of luminosities detected in the most luminous \lya sources at $z=6.6$, with the brightest one being the COLA1 galaxy \citep{Hu2016}. Recently, five additional examples of these so-called ultra-luminous LAEs have been found using Subaru/HSC \citep{Songaila2018}. All these sources seem to have complex line shapes and, in a few cases, even transmitted flux in the blue wing of the line. Their \lya line profile also appears broader than lower luminosity LAEs at the same redshift. 

A quantitative comparison on the width of the \lya line between ultra luminous and average LAEs has been made by \citet{Songaila2018}. We have done a similar analysis on our simulated galaxies by measuring the FWHM of their \lya line profile. We stress that, in our case, we only consider the red part of the line profile (with respect to systemic redshift), because the blue component should be heavily suppressed by the IGM. However, it is interesting to note that one of the ultra luminous galaxies in the \citet{Songaila2018} sample as well as COLA1 both have a detected blue wing. This has been interpreted as the presence of an ionized region around the source, large enough such that \lya photons can redshift away from resonance before reaching the neutral IGM. It is worth mentioning that, in the CW and VW simulations, the most massive galaxy also shows some transmitted flux blueward of the \lya red peak. However, when compared to the systemic redshift of the source, this flux is still transmitted redward of the line center and is due to the presence of another close nearby source.

In the left panel of Fig.~\ref{fig:lya_fwhm}, we show the relation between the width of the \lya line profile, quantified by its FWHM, and the \lya luminosity of the galaxy, measured by \citet{Songaila2018} (based on the Fig.13 in their paper), together with those for the normal and low luminosity LAE samples of \citet{Hu2010} and \citet{Matthee2017}.
The results from our simulated galaxies are shown in the right panel.
Note that many of our simulated galaxies have an apparent \lya luminosity below $10^{42.5}\,{\rm erg\, s^{-1}}$ which corresponds to the detection limit in the \citealt{Songaila2018} observations. For this reason, the bulk of the galaxy population from our simulations is not visible in the Figure and we only focus on the luminous sources to compare with their galaxy sample.

The luminous objects in the CW and VW models have very broad lines with FWHM$\geq 500$ km\,s$^{-1}$. In particular, the most luminous galaxy found in the CW model, which is also the overall most luminous source among all our simulation runs, has FWHM$\sim 700$ km s$^{-1}$, higher than the line width of observed ultra-luminous LAEs of similar apparent \lya luminosity. This is primarily caused by the large outflow velocities present in the CW model, which result in a very broad line emerging from the galaxy according to our simple modeling of the intrinsic line shape emitted from galaxies (Section \ref{sec:methods}). On the other hand, the same galaxy in the VW model has a line width in better agreement with the observed values of $\sim 500$ km s$^{-1}$. This is a good indication that the more physically motivated prescription for galactic outflows used in the VW model has outflow velocities consistent with the ones inferred from diagnostics of the \lya line.

%%%%%%%%%%%%%%%%%%%%%%%%%%%%%%%%%%%%%%%%%%%%%%%%%%%%%%%%%%%%%%%%%%%
\section{Conclusions}
\label{sec:conclusion}
%%%%%%%%%%%%%%%%%%%%%%%%%%%%%%%%%%%%%%%%%%%%%%%%%%%%%%%%%%%%%%%%%%%

We have post-processed the results obtained from high-resolution cosmological zoom-in simulations with a three dimensional Monte-Carlo radiative transfer code to model the \lya properties of simulated galaxies at $z \sim 6.6$. The simulations consist of runs with constrained initial conditions, in which a central $\sim 10^{12} h^{-1}$ M$_\odot$ DM halo is forming, as well as an unconstrained run with the same parent realization of the density field used for comparison. The simulations have been run with three different models for galactic winds. The distribution of HI gas in the simulations have been calculated a posteriori assuming photoionization coming from both the UV background and from galaxies within the simulation box. Self-shielding corrections have been applied in order to include dense, self-shielded gas. Our main findings are as follows:

\begin{itemize}

\item The global distribution and spatial extent of the \lya emission around galaxies is affected significantly by the presence of galactic winds. The model with the strongest feedback from winds (the CW model) has the least extended \lya surface brightness distribution around massive galaxies compared to the model without galactic outflows (NW) or the model with weaker winds (VW). These differences originate not only from the differences in the emergent \lya line profile coming out of the galaxies (which depends on the outflow velocity), but also from the differences in the distribution of neutral gas in the CGM and IGM as a function of galactic wind model. 

\item The luminosities of the most luminous objects match those of recently observed ultra-luminous LAEs at $z \sim 6.6$. In the CR simulations, these galaxies reside in the central, massive DM halo that forms as a result of the imposed constraint and are absent in the simulation corresponding to the normal average field. This suggests that the typical host halo mass of bright LAEs at $z \sim 6.6$, with \lya luminosities in the range $10^{43.5}-10^{44}\,{\rm erg\, s^{-1}}$ is $\sim 10^{12}\, h^{-1}\, \mathrm{M}_\odot$. 

\item The \lya EW distributions of our simulated galaxies is consistent with the one observed for galaxies residing in an overdensity at $z\sim 7$. The EW distributions are similar between CR and UCR simulations, suggesting that the large-scale environment does not affect strongly the \lya EW distributions of LAEs at high-$z$. The VW model, implementing more realistic galactic outflows, is also able to match reasonably well the observed line width of ultra-luminous LAEs at $z\sim 6.6$.

\end{itemize}

%%%%%%%%%%%%%%%%%%%%%%%%%%%%%%%%%%%%%%%%%%%%%%%%%%%%%%%%%%%%%%%%%%%
\section*{Acknowledgements}
%%%%%%%%%%%%%%%%%%%%%%%%%%%%%%%%%%%%%%%%%%%%%%%%%%%%%%%%%%%%%%%%%%%

This work has been partially supported by the  Hubble Theory Grant HST-AR-14584 and by JSPS KAKENHI grant 16H02163 (to I.S.). I.S. is grateful for a generous support from International Joint Research Promotion Program at Osaka University. E.R.D. thanks support from the Collaborative Research Centre 956, sub-project C4, funded by the Deutsche Forschungsgemeinschaft (DFG). At an early stage of the project, R.S. and Z.Z. were supported by NSF grant AST-1208891 and NASA grant NNX14AC89G.
The STScI is operated by the AURA, Inc., under NASA contract NAS5-26555. Numerical simulations have been performed on the University of Kentucky DLX Cluster and using a generous allocation on the XSEDE machines to I.S. The support and resources from the Center for High Performance Computing at the University of Utah are gratefully acknowledged.

%-----------------------
%%%% Bibliography %%%% 
\bibliographystyle{mnras}
\bibliography{lya}
%-----------------------

\appendix

%%%%%%%%%%%%%%%%%%%%%%%%%%%%%%%%%%%%%%%%%%%%%%%%%%%%%%%%%%%%%%%%
\section{Importance of dust on the \lya properties of our simulated galaxies}
\label{app:dust}

\begin{figure}
\includegraphics[width=\linewidth]{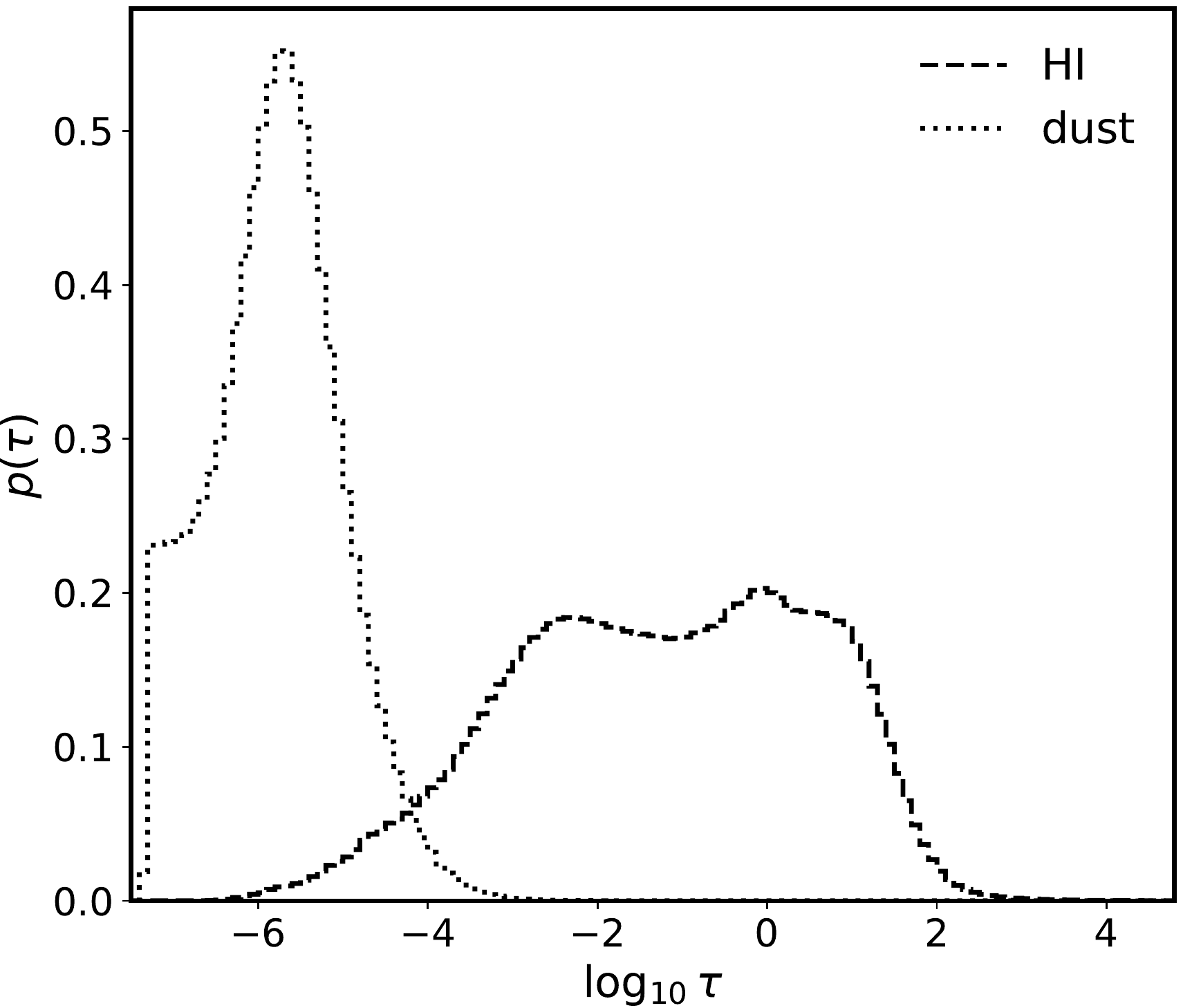}
\caption{Distribution of \lya optical depths due to neutral hydrogen (dashed line) and due to dust (dotted line). In both cases, the optical depth are computed at the \lya line center. The dust optical depth is calculated using the dust radiative transfer model of \citet{Laursen2009b} for \lya photons.}
\label{fig:tau_dust}
\end{figure}

The radiative transfer model we use to predict the \lya properties of our simulated galaxies does not include treatment for the dust. The effect of dust on \lya emission is not a trivial one and depends on the detailed dust properties (e.g. grain size distributions, spatial distribution, clumping factor, among others) in and around galaxies. It can in principle dramatically affect the radiative transfer of \lya photons, since dust can both absorb and scatter \lya photons, thus reducing the overall transmitted flux.  

Although high-$z$ galaxies appear relatively dust free \citep{Bouwens2012,Finkelstein2012}, this might not necessarily be true for the massive galaxies that form in our CR simulations, since, for these objects, the buildup of stellar mass and metallicity is accelerated with respect to galaxies in the field. 

Furthermore, since these galaxies also power strong galactic outflows (in the case of the CW and VW simulation runs), it is likely that the CGM and IGM around them are polluted by dust transported out of the galaxy by these winds. For example, we know that winds transport metals efficiently outside of galaxies and into the IGM, as can be seen on Fig. 9 of S16. In the case of the CW model, metals are found filling most of the IGM within the simulation volume. Since dust is generally associated with metals, one can expect the presence of dust to be ubiquitous in our simulations. On the other hand, another important effect of the winds is to inject thermal energy into the IGM. In the case of the CW model for example, a significant fraction of the IGM gas is at $T\ge 10^4$K, which is sufficient to destroy dust grains and prevent dust formation. In these conditions, it is likely that dust does not have a significant effect on the radiative transfer of \lya photons \emph{in the CGM and IGM}. 

Dust could still be important on ISM scales which we do not resolve in our radiative transfer model since we treat galaxies as point sources for \lya (Section \ref{sec:lyaRT}). This might be an important caveat of our model for the predicted \lya properties of the massive objects in the overdense region, since, as noted before, they should have a non-negligible dust content. 

In order to quantify the effect of dust on the radiative transfer of \lya photons in our simulations, we compare the \lya optical depth due to dust and due to neutral hydrogen in the CW model. To compute the \lya optical depth due to dust, we use the dust attenuation model of \citet{Laursen2009b}, which is explicitly designed to include the effect of dust on the radiative transfer of \lya photons. 
For the HI optical depth, we have included the temperature dependence of the \lya cross-section and used the values of the IGM temperature found in the simulation. In both cases, the optical depths are calculated at the \lya line center. 

The resulting distribution of optical depths for all grid cells are shown in Fig. \ref{fig:tau_dust}. The \lya optical depth due to HI clearly dominates over the dust optical depth. The majority of cells have optical depths due to HI above $\sim 10^{-4}$ while the optical depth due to dust peaks at $\sim 10^{-6}$. From this simple exercise, we conclude that dust should not affect significantly the transfer of \lya photons in the CGM and IGM and, thus, the \lya properties of our simulated galaxies, provided that the \lya line emerging from galaxies is reasonably well captured by our simple model.

\section{Numerical resolution}
\label{app:resolution}

\begin{figure}
\centering
\includegraphics[width=\linewidth]{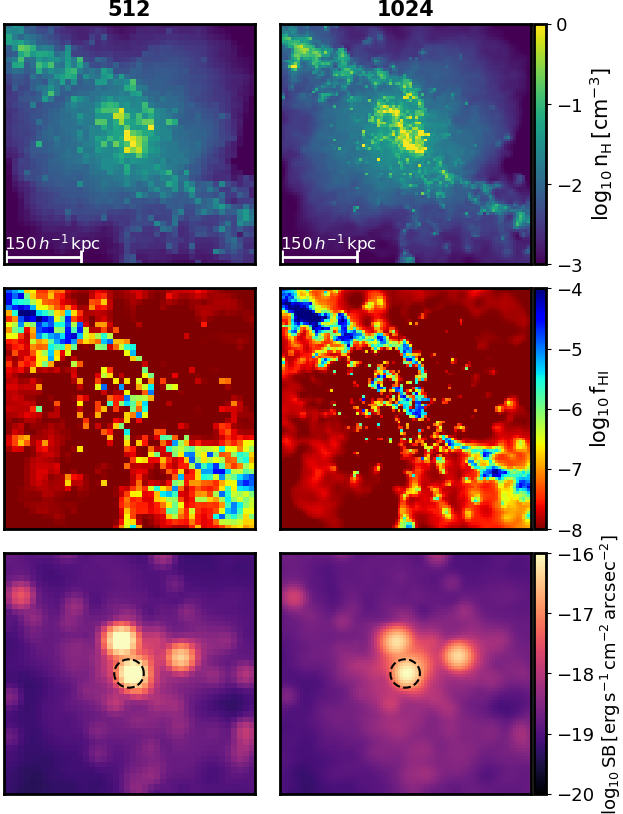}
\caption{Gas distribution around the most massive galaxy. The left column shows the gas distribution with our fiducial grid resolution ($N_{\rm grid}$ = 512) while the right column corresponds to a finer grid resolution ($N_{\rm grid}$ = 1024). The maps show the total gas density (\emph{top}), neutral fraction (\emph{middle}) and \lya surface brightness (\emph{bottom}) in a $0.5\,h^{-1}\,\mathrm{Mpc}$ region around the galaxy.}
\label{fig:maps_gal0_512_1024}
\end{figure}

\begin{figure}
\centering
\includegraphics[width=0.8\linewidth]{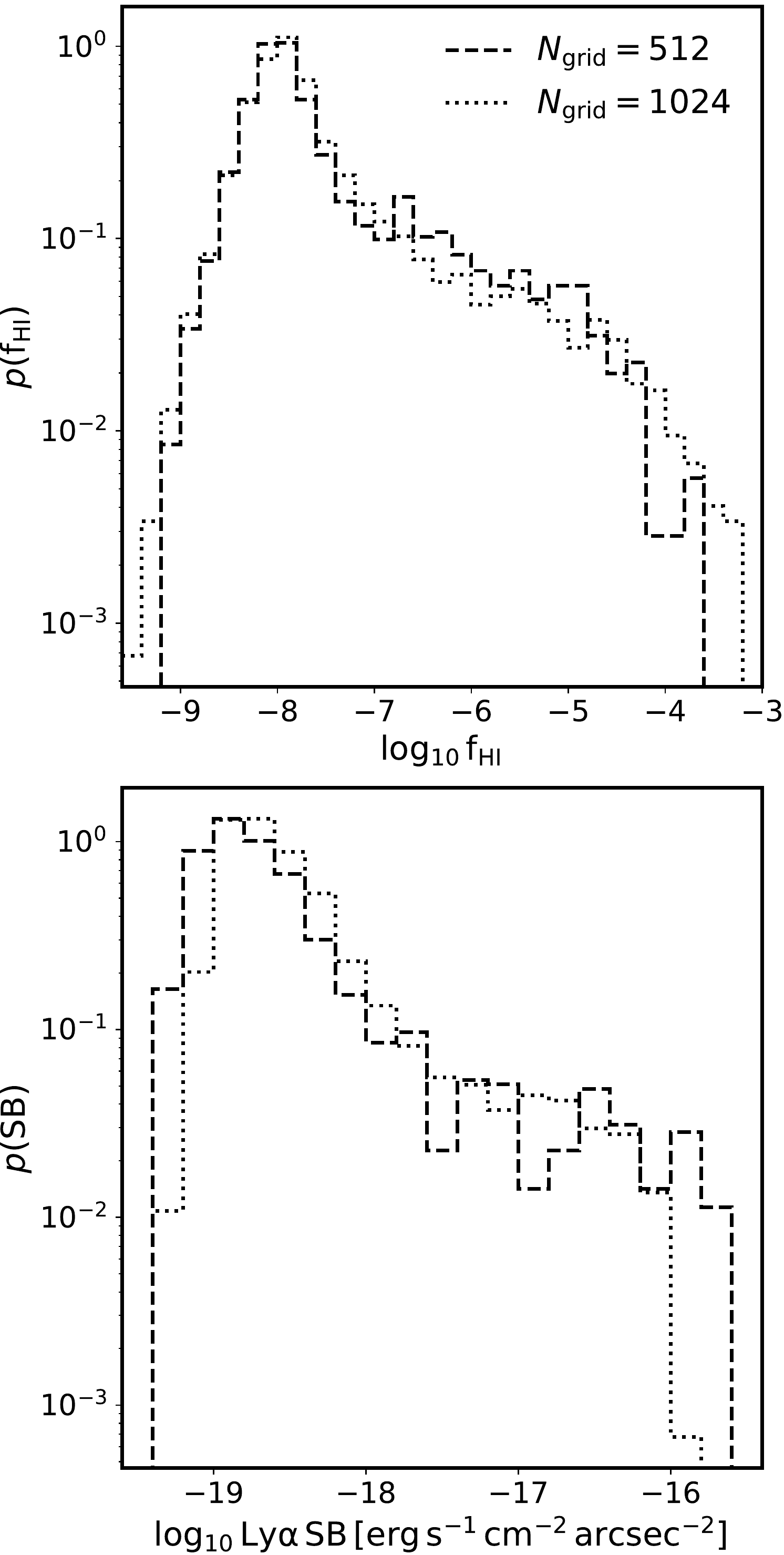}
\caption{Distributions of pixel values in the neutral fraction (\emph{top}) and \lya surface brightness  (\emph{bottom}) maps shown in Fig. \ref{fig:maps_gal0_512_1024}. In each panel, the dashed line indicates the results for $N_{\rm grid} = 512$ while the dotted line corresponds to the case of $N_{\rm grid} = 1024$.}
\label{fig:distribs_512_1024}
\end{figure}

Due to the resonant scattering of \lya photons, their radiative transfer depend sensitively on the spatial distribution of neutral gas in the CGM and IGM. In particular, the presence of self-shielded absorbers on small-scales (e.g., Lyman-limit systems) can significantly alter the \lya transmission and resulting \lya properties of LAEs. For this reason, the \lya properties of our simulated galaxies should in principle be sensitive to the spatial resolution of the grid we use to perform the \lya radiative transfer calculations (Section \ref{sec:sph_to_grid}). If the resolution of the grid is insufficient to properly resolve these self-shielded systems in the CGM and IGM, our models could be overestimating the apparent \lya luminosity of galaxies. As mentioned in Section \ref{sec:methods}, the results presented in the paper are obtained using a grid of size $N_{\rm grid} = 512$ to interpolate the gas properties from the SPH data in the simulations. Here we investigate whether using a finer grid to better resolve the small scale structures in the neutral gas distributions affects our results on the \lya properties of galaxies in the CW model. 

Fig. \ref{fig:maps_gal0_512_1024} shows the gas density, HI fraction and \lya surface brightness maps in a $0.5\,h^{-1}\, \mathrm{Mpc}$ region centered around the most massive galaxy in the CW model. The left panels show the results for our fiducial grid with $N_{\rm grid} = 512$, while the right panels correspond to the case for $N_{\rm grid} = 1024$. These maps show that, qualitatively, the spatial distributions of dense, neutral regions are essentially the same between the two cases. In both cases, these regions trace the dense filaments as well as the CGM gas close to the central galaxy. Since the neutral gas distributions share similar features, it is not surprising to find that the spatial extent and absolute values in \lya surface brightness are also qualitatively the same (besides the obviously higher spatial resolution in the $N_{\rm grid} = 1024$ case).  

For a more quantitative comparison, we have computed the distribution of neutral fraction, $\mathrm{f}_{\rm HI}$, and \lya surface brightness values from all pixels within the maps shown in Fig. \ref{fig:maps_gal0_512_1024}. These distributions are shown in Fig. \ref{fig:distribs_512_1024}. We find that the distributions of neutral fractions are in good agreement between the two cases, $N_{\rm grid} = 512$ and $N_{\rm grid} = 1024$. In particular, both distributions have similar tails extending towards large neutral fractions, which indicates that our fiducial model with $N_{\rm grid} = 512$ should be sufficient to resolve dense structures in the neutral gas. This is reflected in the \lya surface brightness distributions which are also in good agreement in the range $\sim 10^{-19}-10^{-16}\, {\rm erg\, s^{-1}\, cm^{-2}\,arcsec^{-2}}$. Based on these results, we conclude that $N_{\rm grid} = 512$ is sufficient to properly resolve the neutral gas distributions in the CGM and IGM, thus ensuring that the \lya properties presented in the paper should not suffer from numerical resolution issues.

\end{document}